\documentclass[prb,twocolumn,longbibliography,amsmath,amssymb,floatfix,]{revtex4-2}

\usepackage{graphicx}
\usepackage{color}
\usepackage{pst-node}
\usepackage{bbold}
\usepackage[unicode]{hyperref}
\hypersetup{unicode=true, colorlinks=true, linkcolor=black, citecolor=black}

\def \br{{\bf r}}

\def \bk{{\bf k}}
\def \bp{{\bf p}}

\def \b1{{\bf 1}}

\newcommand{\bea}{\begin{eqnarray}}

\newcommand{\eea}{\end{eqnarray}}

\newcommand{\beq}{\begin{equation}}

\newcommand{\eeq}{\end{equation}}

\def \br{{\bf r}}

\def \bk{{\bf k}}
\def \bp{{\bf p}}

\def \be{{\bf e}}

\def \b0{{\bf 0}}

\usepackage{ulem}

\newlength\figurewidth
\newlength\figurefullwidth
\begin{document}

\title{Time crystals in cavity-BEC systems}

\begin{abstract}
The understanding of light-induced dynamical states continues to be a challenging and fruitful pursuit of science.
 This pursuit is supported by quantum simulation  of dynamical phenomena, e.g., in ultracold atom systems.
 Typically, ultracold atom dynamics are read out destructively, via time-of-flight imaging, limiting a detailed analysis.
 However, atom-cavity systems provide a real-time readout of the photonic state via photon emission from the cavity, making the system ideally suited for the simulation of dynamical phenomena.
 Here, we review three distinct time crystalline states, predicted and realized in a cavity-BEC system.
 We give an example for each of them, based on  minimal few-mode models. 
 We characterize the time crystalline states via correlation functions of the cavity mode, and characteristic momentum modes of the condensate.
 This supports a clear distinction between these time crystals. 
More generally, the sequence of studies reviewed  here, serves as a blueprint for setting up minimal models and their characterization, for dynamical phenomena. 
\end{abstract}  

\author{Jayson G. Cosme}
\affiliation{National Institute of Physics, University of the Philippines, Diliman, Quezon City 1101, Philippines}

\author{Ludwig Mathey}
\affiliation{Center for Optical Quantum Technologies and Institute for Quantum Physics, Universit\"at Hamburg, 22761 Hamburg, Germany}
\affiliation{The Hamburg Center for Ultrafast Imaging, Luruper Chaussee 149, 22761 Hamburg, Germany}

\date{\today}

\maketitle

\tableofcontents

\section{Introduction}

Time crystals are non-equilibrium states found in both classical and quantum many-body systems defined by their spontaneous breaking of time-translation symmetry \cite{Sacha2018,Else2020,zaletel2023}. This temporal-symmetry breaking manifests itself in the dynamics of an observable $\hat{O}$ exhibiting an emergent periodicity in time, $T_\mathrm{TC}$. That is, an observable, such as the two-time correlation function, follows $\langle \hat{O}(t) \rangle = \langle \hat{O}(t+T_\mathrm{TC}) \rangle$, wherein $T_\mathrm{TC}$ is strictly not equal to any time scales imposed by the dynamical equations governing the system's evolution. In addition to emergent patterns in time, time crystals are also defined by their robustness against fluctuations and perturbation, similar to the rigidity of conventional crystals or solids. Furthermore, as a genuine many-body phase of matter, interaction between the constituents of the system must play a crucial role in the emergence of a time crystal. The defining features of most time crystals reported in the literature are:
\begin{enumerate}
\item Time-translation symmetry breaking 
\item Interaction-induced or many-body nature
\item Robustness against perturbations
\end{enumerate}

Originally envisaged by Frank Wilczek \cite{Wilczek2012}, the existence of time crystals in equilibrium scenarios is challenged by a series of no-go theorems, which can be understood as energy conservation preventing a system from spontaneously oscillating in the absence of any external force \cite{Noz2013,Bruno2013,Watanabe2015}. To circumvent the no-go theorems, the search for time crystals has shifted focus to non-equilibrium platforms, such as periodically driven systems and dissipative or open many-body systems. For periodically driven many-body systems, the simplest time crystal exhibits period doubling in the dynamics of its observables, $\langle \hat{O}(t) \rangle = \langle \hat{O}(t+2T_D) \rangle$, despite the Hamiltonian possessing a temporal periodicity of $T_D$, i.e., $\hat{H}(t) = \hat{H}(t+T_D)$. Consequently, the system exhibits a subharmonic response, oscillating at a fraction of the driving frequency and thus breaking the discrete time-translation symmetry, a hallmark of discrete time crystals. More generally, discrete time crystals arise whenever a system exhibits subharmonic dynamics at an integer multiple of the driving period, $\langle \hat{O}(t) \rangle = \langle \hat{O}(t+nT_D) \rangle, n \in \mathbb{Z}, n>1$. However, there is a major hurdle for stabilising time crystals in a generic closed many-body system: periodic driving results in heating and loss of memory, leading to the eventual decay or melting of the time crystal. Thus, earlier realisations of discrete time crystals in periodically driven many-body systems have focused on delaying thermalisation using disorder and specific forms of interactions \cite{Else2016,Yao2017,Choi2017,Zhang2017}.

    \begin{figure*}
\includegraphics[trim={0.1cm 2.0cm 2.0cm 0.1cm},clip,width=0.5\textwidth]{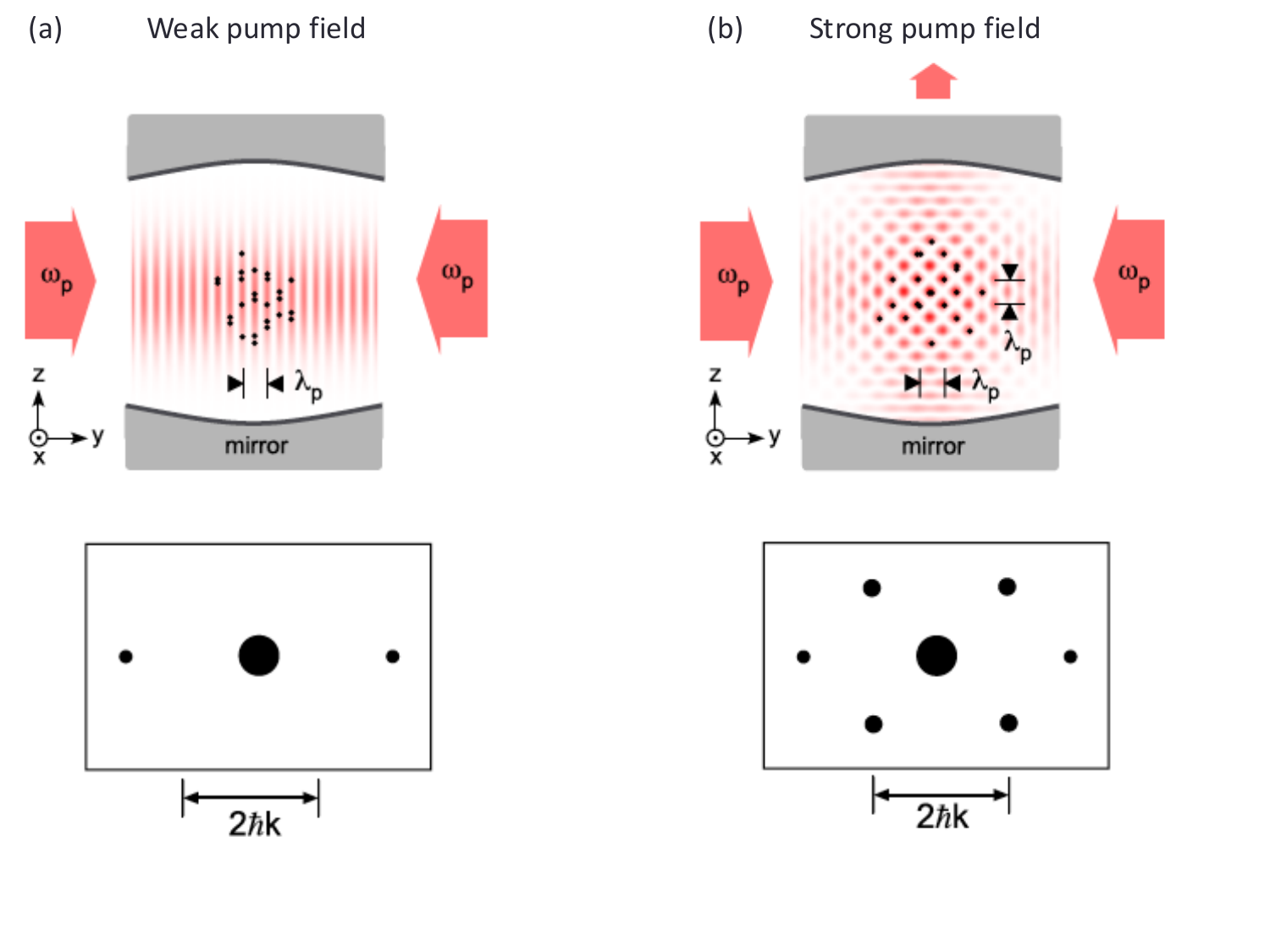}
\caption{As the platform for creating time crystals, we consider a cavity-BEC system that is pumped from the transverse direction. 
  For weak pump field, panel (a), the standing field of the retroreflected pump beam creates a weak optical lattice. In momentum space, shown in the lower panel,
  two side peaks are generated at $\pm 2 k \be_{z}$, in addition to the condensate peak at $\bk = 0$.
    The cavity is essentially silent.
    For a strong pump field, the system forms a superradiant state. The atoms form a checkerboard lattice, generating additional side peaks at $\pm k \be_{y} \pm k \be_{z}$, shown in the lower panel.
   The cavity mode is activated, leading to photoemission through the semitransparent mirror. 
    The transition from the weak-pump to the strong-pump state is a Dicke transition. 
  \label{cavbec}}
\end{figure*} 

In dissipative platforms, where a system interacts with an environment, the melting of time crystals can be inhibited by an appropriately engineered dissipation channel. Here, dissipation plays a constructive role in stabilising the time crystal as it extracts energy from the system, leading to the suppression of heating. While the concept of dissipation-stabilised or dissipative time crystals was first introduced in Ref.~\cite{Else2017}, a concrete proposal for a period-doubling time crystal in a dissipative system, specifically the open Dicke model, was first put forth in Ref.~\cite{Gong2018}. It was suggested that circuit- or cavity-QED setups can be used to realise these Dicke time crystals \cite{Zhu2019,Simon2024}. This is precisely the type of dissipative time crystal first realised in a cavity-BEC system \cite{Kessler2021}. Further exploration has since revealed that cavity-BEC systems host a rich and diverse ecosystem of time crystals beyond the Dicke time crystals, including incommensurate and continuous time crystals, as predicted in Refs.~\cite{Cosme2019,Kessler2019,Kessler2020,Skulte2021} and experimentally observed in Refs.~\cite{Kongkhambut2021,Kongkhambut2022,Kongkhambut2024,Skulte2024,Cosme2025}.

To understand how the cavity-BEC system emulates the paradigmatic discrete time crystal of the Dicke model, we briefly review the fundamental connection between these two systems. In doing so, we can highlight how the complexity and multimode nature of the cavity-BEC setup, extending beyond the standard Dicke model, enables it to support a  diverse collection of time crystals. We consider a specific configuration of the atom-cavity platform consisting of a Bose-Einstein condensate inside an optical resonator with transverse pumping as depicted in Fig.~\ref{cavbec}. 
This setup has been shown to exhibit a phase transition between two static phases, which are a homogeneous BEC and a self-organised chequerboard density wave (DW) phase, triggered by varying the intensity of the transverse pump field \cite{Baumann2010,Klinder2015}. Through a series of transformations and a two-mode approximation of the matter sector, the Hamiltonian for the cavity-BEC system can be mapped onto that of the Dicke model \cite{Mivehvar2021}. The standard open Dicke model describes an ensemble of qubits or spin-1/2 particles interacting with a single quantised mode of light experiencing single-photon dissipation \cite{DickeModel} and is known to host a normal-superradiant phase transition. Then, the BEC-DW  transition corresponds to the normal-superradiant phase transition in the Dicke model, establishing the transversely pumped cavity-BEC setup as a versatile quantum simulator for the Dicke model.

For larger pump intensities, the two-mode approximation of the cavity-BEC breaks down as higher momentum modes become significantly occupied, eventually leading to a departure from the physics of the standard Dicke model. This inherent complexity of the cavity-BEC systems ultimately allows for more exotic time crystals to emerge. Indeed, incommensurate time crystals have been predicted to emerge in shaken cavity-BEC systems due to the resonant excitation of a momentum mode absent in the static case \cite{Cosme2019,Skulte2021}. An incommensurate time crystal is characterised by an irrational ratio between the time-crystal period and the driving period, $T_\mathrm{TC}/T_D \notin \mathbb{Q}$. 
Furthermore, time crystals have been observed in cavity-BEC systems even in the absence of periodic driving leading to the emergence of so-called continuous time crystals \cite{Kongkhambut2022}, the type of time crystal closest to Wilczek's original vision. These continuous time crytals, which can be understood as limit cycles in the language of nonlinear dynamics \cite{Skulte2024}, arise as an instability of the superradiant phase at large pump intensities. This instability is attributed to the occupation of a third momentum mode beyond the two-mode description of the superradiant state. Both incommensurate and continuous time crystals in cavity-BEC systems rely on a third atomic mode becoming relevant \cite{Skulte2021,Skulte2024}, signalling physics beyond the two-level Dicke model, as will be discussed later in detail. 

For a detailed discussion on the subtleties of the spontaneous breaking of time-translation symmetry in time crystals for closed and open systems, together with a selection of specific realisations, we refer the interested reader to the review articles \cite{Sacha2018,Else2020,zaletel2023} and references therein. In the following, we focus on dissipative time crystals in cavity-BEC systems. Specifically, we highlight three types of dissipative time crystals found in the good-cavity regime, where the photons evolve on the same time scale as the atoms: (i) discrete \cite{Kessler2021}, (ii) incommensurate \cite{Cosme2019,Skulte2021,Kongkhambut2021}, and (iii) continuous \cite{Kessler2019,Kongkhambut2022}. The transition between discrete and continuous time crystals via resonant driving \cite{Kessler2020,Kongkhambut2024}, as well as the instability of a continuous time crystal toward a continuous time quasicrystal \cite{Cosme2025}, has also been investigated in cavity-BEC platforms.

\section{Cavity-BEC systems}
In Fig. \ref{cavbec}, we show a sketch of the system that we use to create time crystalline states. As mentioned above, we consider a Bose-Einstein condensate loaded in a high-finesse cavity. The condensate is pumped from the transverse direction, with a retro-reflected beam. The frequency of the pump beam is far detuned from an internal transition of the atom, and therefore operates in the dispersive regime. 
 Thus, the atoms only occupy a single internal state, while their motional states respond and interact with the pump and cavity fields.

 If the pump intensity is below a critical value, the retro-reflected pump beam acts like a weak optical lattice beam, as depicted in Fig.~\ref{cavbec}(a). The condensate displays two momentum side peaks, predominantly, given by $\pm 2 \hbar k \be_{z}$, where $k$ is the wavevector of the pump laser. 
 The cavity is essentially silent in this state.
  If the pump intensity is above a critical intensity,  as depicted in Fig.~\ref{cavbec}(b), the pump photons are  scattered into the cavity mode. This scattering process generates atoms in the momentum states $\pm k \be_{y}\pm k \be_{z}$  in the $y-z$ plane, i.e. the plane composed of the pump direction and the cavity direction.
  This in turn supports Bragg scattering in and out of the cavity, resulting in a self-organized checkerboard density pattern, which
 breaks  the $\mathbb{Z}_{2}$ symmetry associated with the checkerboard order.
   As we describe below, this superradiant transition can be captured in an effective two-mode model, and closely resembles a Dicke transition.

\section{Time crystals in cavity-BEC systems}
\begin{figure*}
\includegraphics[trim={0.1cm 1.8cm 1.0cm 0.1cm},clip,width=0.55\textwidth]{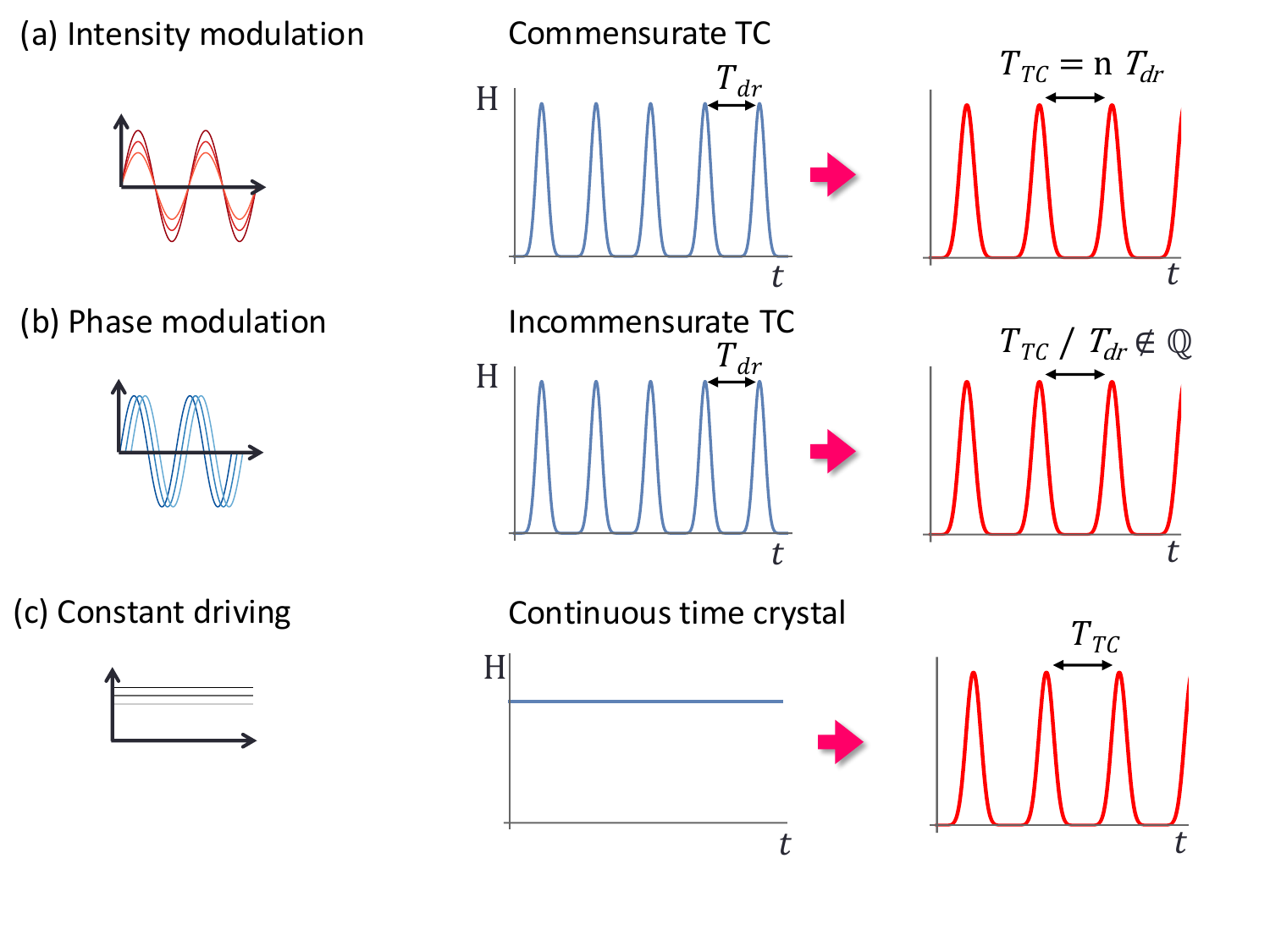}
\caption{We discuss three time crystalline states, emerging in a cavity-BEC system. By modulating the intensity of the pump beam, we induce a commensurate, discrete time crystal. By modulating the phase of the pump beam, we induce an incommensurate time crystal. For continuous driving, we induce a continuous time crystal.
 These time crystalline states are defined via the time-dependent correlation functions of the cavity mode and specific atomic modes, as we discuss in this review. 
   \label{ovtc}}
\end{figure*}

 In this article, we describe three time crystalline states that emerge in a cavity-BEC system.
 In Fig. \ref{ovtc} we display these states as a sketch.
 In panel (a), we consider intensity modulation of the pump laser. Below we give an example for a regime in which this type of periodic driving results in a commensurate, discrete time crystal. The Hamiltonian is periodic in time with the time period $T_{dr}$ of the intensity modulation, thus it breaks time translational symmetry explicitly. More precisely, the continuous time translational symmetry, associated with a time-constant Hamiltonian, is reduced to a discrete, lattice-like symmetry, in which the Hamiltonian is invariant under the discrete set of translations given by $\{ n T_{dr}\}$.
  The emergent steady state of this system breaks this remaining discrete symmetry. Specifically, it is a period-doubling state, in which the system undergoes oscillations between the two superradiant states of the equilibrium system \cite{Kessler2021}.

  In panel (b), we display a drive due to a phase modulation of the pump laser. This type of periodic driving generates an incommensurate time crystal, for which we give an example below. Here, the periodic driving reduces the time translational symmetry to a discrete lattice symmetry, as it did for intensity modulation, however, the emergent steady state displays an oscillation period that is not commensurate to the driving period. No integer or rational ratio of small numbers can be identified, it is thus an incommensurate time crystal. This terminology is analogous to charge density waves occurring in solid state systems, in which electrons order in density waves that are commensurate or incommensurate to the spatial lattice.
   As we describe below, the two-mode description that we have mentioned above, which serves as a minimal model for the Dicke transition, has to be expanded to a three-mode description \cite{Cosme2019,Skulte2021,Kongkhambut2021}.

  In panel (c), we consider a dynamical regime of the system with constant driving. This expands the dynamical regimes described above, beyond the ordered and disordered state of the Dicke transition.
   For sufficiently weak detuning, the above mentioned two-mode approximation is not sufficient, qualitatively, but needs to be expanded to a three-mode model, however, with a different third mode as for the case of phase modulation.
   The resulting Hamiltonian is constant in the rotating frame, therefore the time-translational symmetry is continuous.
    This continuous symmetry is spontaneously broken by the time crystalline state, which constitutes a continuous time crystal, in a dissipative system \cite{Kessler2019,Kongkhambut2022}.
    
    In the following, we give an example for each of these states, in a tutorial-like fashion. 
    For each, we work in a minimal few-mode model. A more extensive discussion of the full dynamical phase diagrams and their properties, is given in the references.

\section{Two-mode model}
 As a starting point, 
  we derive a minimal model for the description of the Dicke ordering transition by
  utilizing two momentum components of the condensate, and additionally a single cavity mode. 
 First, we consider the full multi-mode Hamiltonian $H = H_{C} + H_{A} + H_{AC}$, with the cavity Hamiltonian given by
\bea
\frac{H_{C}}{\hbar} &=& - \delta_{C} a^{\dagger} a.
\eea 
 $\delta_{C}$ is the detuning between the pump beam and the atomic transition, i.e. $\delta_{C} = \omega_{P} - \omega_{C}$, where $\omega_{P}$ is the frequency of the pump beam, and $\omega_{C}$ is the cavity frequency.  $a$ is the photon operator. 
 The atomic Hamiltonian $H_{A}$ is given by
\bea
 \frac{H_{A}}{\hbar}&=& \int d^{2}r \Psi^{\dagger}(\br) \Big( \!\! -\frac{\hbar \nabla^{2}}{2 m} \!
 - \omega_{rec} \epsilon_{p} \cos^{2}(k y) \!\Big) \Psi(\br).
\eea
$m$ is the atomic mass, $\omega_{rec}$ is the recoil frequency, i.e. $\omega_{rec} = \hbar k^{2}/(2m)$, and $\epsilon_{p}$ is the unitless pump intensity. 
 As described above, the $y$-direction is the pump direction, see Fig. \ref{cavbec}. 
  Therefore, the retroreflected pump laser crates an optical lattice along the $y$-direction, given by the potential $- \omega_{rec} \epsilon_{p} \cos^{2}(k y)$. 
 The interaction $H_{AC}$ between the atoms and the cavity mode is given by
\bea
 \frac{H_{AC}}{\hbar}&=& \int d^{2}r \Psi^{\dagger}(\br) \Big( U_{0} a^{\dagger} a \cos^{2}(k z)\nonumber\\
&&  +  2 g \sqrt{\epsilon_{p}} \cos(k y)\cos(k z)(a^{\dagger} + a)  \Big) \Psi(\br)
\eea
  $U_{0}$ is the light shift per photon, and $g$ is given by
  $g = \text{sgn}(U_{0}) \sqrt{\omega_{rec} |U_{0}| }/2$.
 The coupling term that is proportional to $\sim (a^{\dagger} + a)$ is the origin of the Dicke transition.
  In the ordered state, $\langle (a^{\dagger} + a)\rangle$ acquires a positive or negative expectation value, corresponding to a spontaneous $\mathbb{Z}_{2}$ symmetry breaking.
   In these ordered states, the atoms move in a light-induced potential of the form $\cos(k y)\cos(k z)$. Whether the prefactor of this potential is positive or negative, depends on which symmetry-broken state emerges in the cavity mode.
   Therefore, which sublattice of potential minima that the atoms order in, is tied to the broken symmetry in the cavity mode.

 Written in momentum representation via $\Psi(\br) = \frac{1}{\sqrt{V}} \sum_{\bp} \exp(i \bp \cdot \br) b_{\bp}$, the atomic Hamiltonian $H_{A}$ takes the form
 \bea
\frac{H_{A}}{\hbar}&=& \sum_{\bp} \Big( \omega_{p}  - \frac{\omega_{rec} \epsilon_{p}}{2} \Big) b_{\bp}^{\dagger} b_{\bp}\nonumber\\
&& - \frac{\omega_{rec} \epsilon_{p}}{4}  \sum_{\bp} ( b_{\bp}^{\dagger} b_{\bp+2k\be_{y}} + b_{\bp}^{\dagger} b_{\bp-2k\be_{y}})
\eea
 with the atomic dispersion $\omega_{p} = \hbar p^{2}/(2 m)$.
The interaction Hamiltonian $H_{AC}$ takes the form
\bea
 \frac{H_{AC}}{\hbar} &=& \frac{U_{0}}{2} a^{\dagger} a \sum_{\bp} b_{\bp}^{\dagger} b_{\bp}\nonumber\\
 &&+ \frac{U_{0}}{4} a^{\dagger} a \sum_{\bp} b_{\bp}^{\dagger} (b_{\bp +2 k \be_{z}} + b_{\bp - 2 k \be_{z}})\nonumber\\
&& + \frac{\text{sgn}(U_{0}) \sqrt{\omega_{rec} |U_{0}| \epsilon_{p}}}{4} (a^{\dagger} + a)\nonumber\\ 
&& \sum_{\bp} b_{\bp}^{\dagger} (b_{\bp + k\be_{y} + k \be_{z}} + b_{\bp + k\be_{y} - k \be_{z}}\nonumber\\
&& + b_{\bp - k\be_{y} + k \be_{z}} + b_{\bp - k\be_{y} - k \be_{z}})
\eea
 We take the momentum space representation of $H_{A} + H_{C} + H_{AC}$ as the starting point for the few-mode expansions throughout this review.

        \begin{figure}
\includegraphics[width=0.5\textwidth]{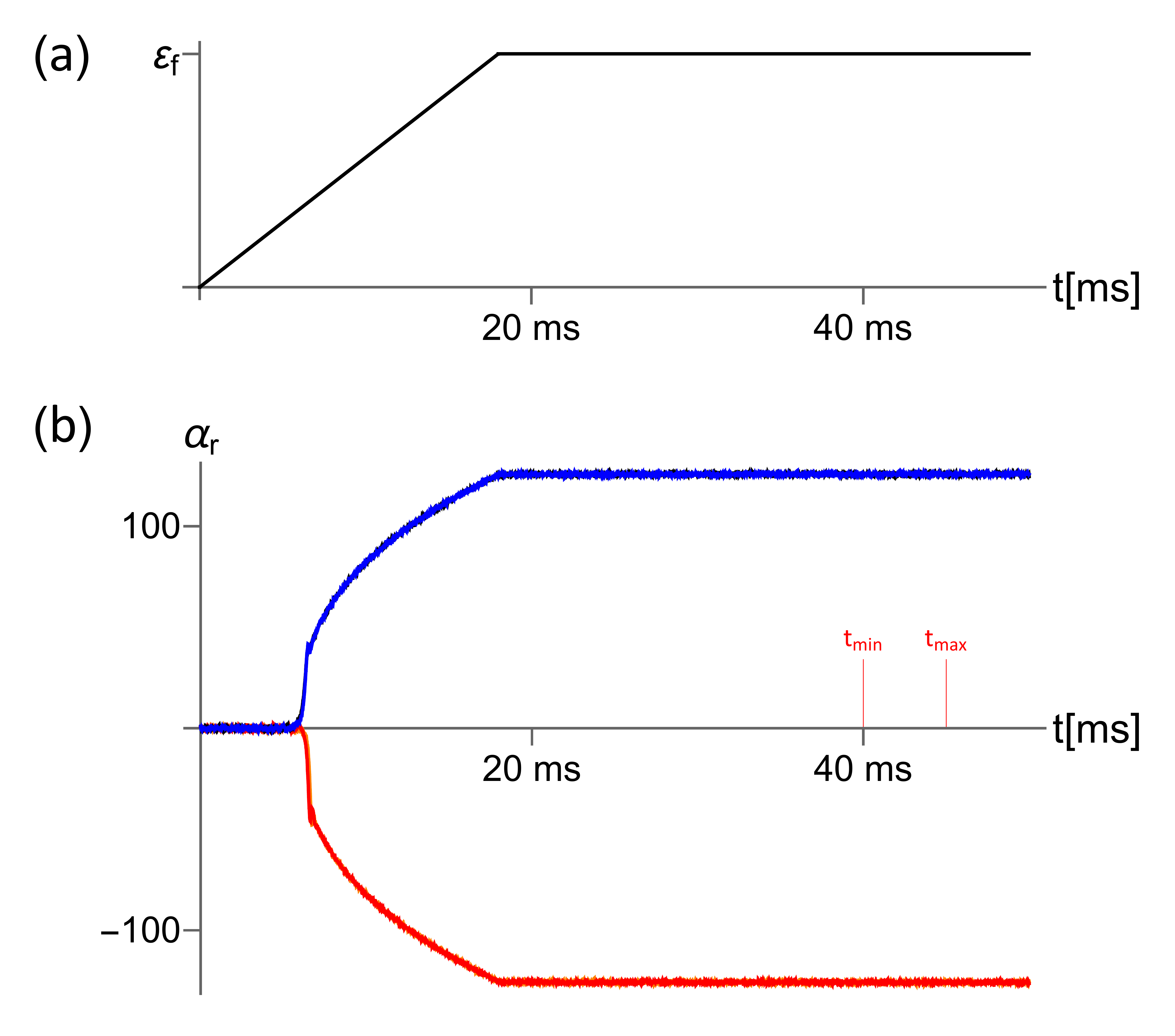}
\caption{To illustrate the dynamical evolution in terms of trajectories, we show the linear ramp into the Dicke ordered state. 
  In panel (a) we show the pump protocol, which is linear up to a time $t_{ramp} = 18$ ms, and constant after that at $\epsilon_{p}(t) = \epsilon_{f} = 3.3$.
  In panel (b) we show several trajectories $\alpha_{r} (t)$, generated by the Heisenberg-Langevin equations Eqs. \ref{dickeeqalpha} --\ref{dickeeqbetae}.
    They display dynamical $\mathbb{Z}_{2}$ symmetry breaking, and then settle in one of the two ordered states.
 \label{trajsuprad}}
\end{figure}

\subsection{Two-mode expansion}
 As a minimal model for the approximate realization of the Dicke transition, we consider the two-mode approximation of the atomic sector, given by the modes
 \bea
 b_{g} &=& b_{\bk=0}\\
 b_{e} &=& \frac{1}{2}( b_{k\be_{y} + k \be_{z}}
  + b_{ k\be_{y} - k \be_{z}}\nonumber\\
  && + b_{ -k\be_{y} + k \be_{z}} + b_{ - k\be_{y} - k \be_{z}} )
 \eea
  We note that the momenta contained in $b_{e}$ are the ones of the form $\cos(k y)\cos(k z)$, discussed above.
  We assume that all other modes have zero occupation, and that the total atom number $N_{a}$ is constant, so that $b_{g}^{\dagger}b_{g} + b_{e}^{\dagger} b_{e} = N_{a}$.
 The Hamiltonian is then given by
  \bea
 \frac{H_{2m}}{\hbar} &=& -\delta_{eff} a^{\dagger} a + 2 \omega_{B} b_{e}^{\dagger} b_{e} + \frac{U_{0}}{4} a^{\dagger} a b_{e}^{\dagger} b_{e}\nonumber\\
 && + g \sqrt{\epsilon_{p}} (a^{\dagger} + a)  (b_{g}^{\dagger} b_{e} + b_{e}^{\dagger} b_{g} ) \label{H2m}
 \eea
with $\delta_{eff} = \delta_{C} - U_{0}N_{a}/2$, and $\omega_{B} = \omega_{B}(\epsilon_{p}) = \omega_{rec} - \omega_{rec}\epsilon_{p}/8$.

       \begin{figure}
\includegraphics[width=0.5\textwidth]{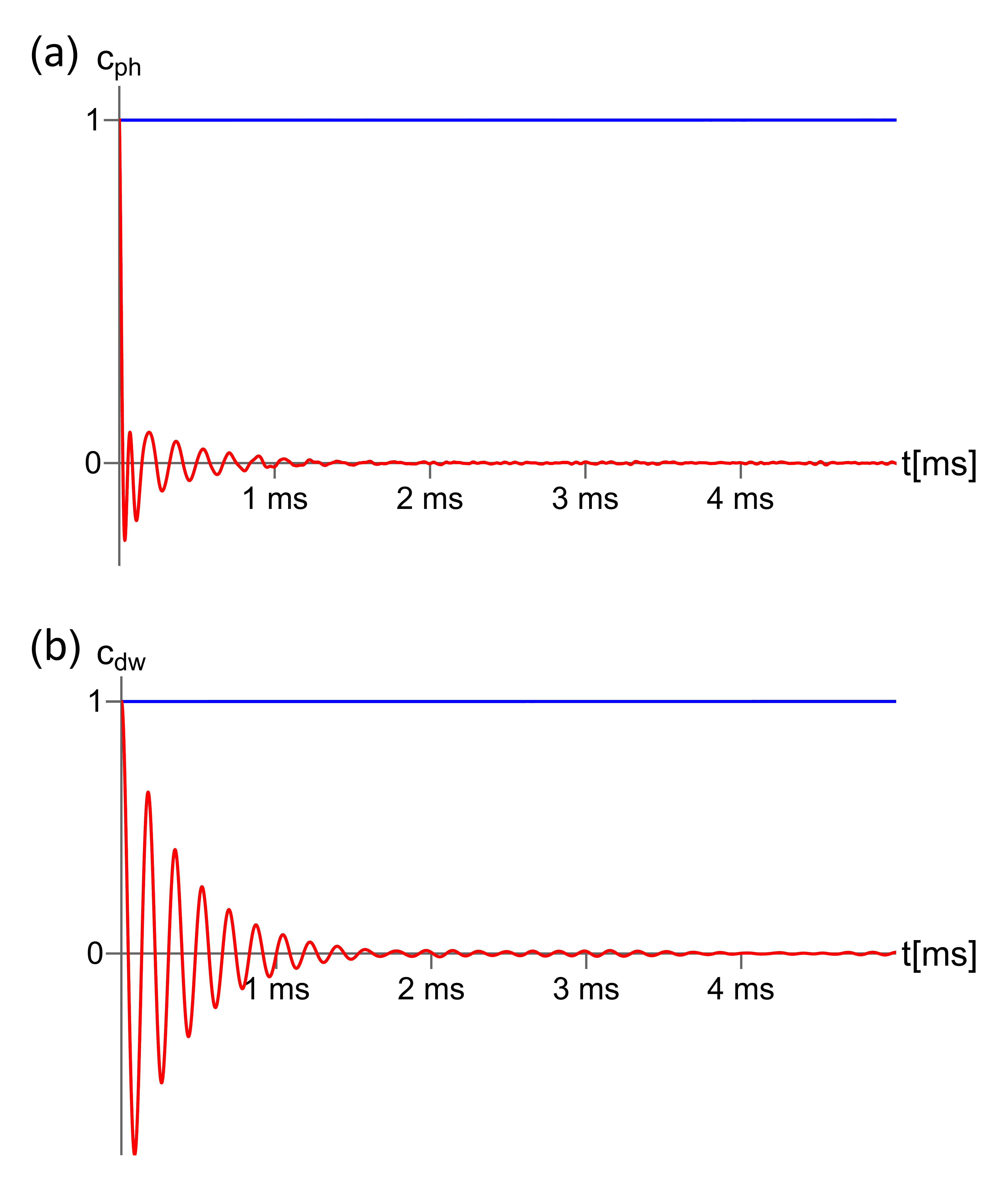}
\caption{In panel (a) we show the real part of the correlation function of the photon mode $c_{ph}(t)$ for the ordered (blue) and disordered state (red).  
 The ordered state  has been generated with $\epsilon_{f} = 3.3$, based on the protocol shown in Fig. \ref{trajsuprad} (a),  
  and the disordered state has been generated with $\epsilon_{f} =0.3$. 
 In panel (b) we show the correlation function of the checkerboard density order $c_{dw}(t)$ for the ordered (blue) and disordered state (red).  
  The fast decay to zero for the disordered state, or to a non-zero value for long times, for the ordered state, indicates temporal short-range or long-range order, respectively. 
\label{figsuprad}}
\end{figure}

\subsection{Semiclassical dynamics}
We approximate the dynamics of the model in Eq.  \ref{H2m} via a semiclassical Heisenberg-Langevin set of equations:
\bea
\frac{d \alpha}{d t} &=&  i \delta_{eff} \alpha - \kappa \alpha - i g \sqrt{\epsilon_{p}} (\beta_{g}^{*} \beta_{e} + \beta_{e}^{*} \beta_{g})\nonumber\label{dickeeqalpha}\\
&& -i \frac{U_{0}}{4} \beta_{e}^{\dagger}\beta_{e} \alpha + \xi\\
\frac{d \beta_{g}}{d t} &=& - i g \sqrt{\epsilon_{p}} (\alpha^{*} + \alpha) \beta_{e}\label{dickeeqbetag}\\
\frac{d \beta_{e}}{d t} &=& - 2 i \omega_{B} \beta_{e} - i g \sqrt{\epsilon_{p}} (\alpha^{*} + \alpha) \beta_{g}\nonumber\\
&& -i \frac{U_{0}}{4} \alpha^{\dagger} \alpha \beta_{e}\label{dickeeqbetae}
\eea
 We initialize these equations with $\alpha(0) = z_{a}$, where $z_{a}$ is a complex, random number variable drawn from the distribution $\frac{2}{\pi} \exp(-2 |z_{a}|^{2})$, i.e. the Wigner distribution of the vacuum state.
 Similarly, we initialize  $\beta_{e}(0) = z_{e}$ from the distribution $\frac{2}{\pi}  \exp(-2 |z_{e}|^{2})$.
   The ground variable $\beta_{g}$ is sampled via $\beta_{g}(0) = \sqrt{N_{a}+1/2} \exp(i\phi)$, where $\phi$ is drawn from a homogeneous distribution on $[0, 2\pi]$.
     This initialization is an approximate representation of the vacuum state for $a$ and $b_{e}$, and a condensate state for $b_{g}$.
      The equation of motion of the cavity mode contains the photon decay $- \kappa \alpha$, with the decay rate $\kappa$.
      Additionally, the back-action of the photon decay is included via the noise term $\xi(t)$, with $\langle \xi(t_{1}) \xi(t_{2})\rangle =\kappa\delta(t_{1} - t_{2} )$.
   We integrate these stochastic differential equations numerically with a Heun algorithm of second order.

\subsection{Correlation functions}
To characterize the ordered state, we introduce two correlation functions. For the cavity mode, we define
\bea
c_{ph}(t)&=& \frac{1}{N_{ph}} \langle a^{\dagger}(t_{0}+t) a(t_{0})\rangle_{t_{0}}\label{corra}
\eea
It describes the temporal coherence of the cavity mode.  In this expression, we average over a time interval of $t_{0} \in [t_{min}, t_{max}]$. 
 The normalization constant $N_{ph}$ is chosen such that $c_{ph}(t=0) = 1$.

To capture the Dicke ordered state in the atomic sector, we define the density component:
\bea
\rho_{dw} &=& \rho_{k \be_{y} +k \be_{z}} + \rho_{k \be_{y} -k \be_{z}}\nonumber\\
&&+ \rho_{-k \be_{y} +k \be_{z}} + \rho_{-k \be_{y} -k \be_{z}}
\eea
 with $\rho_{k} = \sum_{p} b_{p}^{\dagger} b_{p+k}$.
  Written in the two-mode expansion, we have
  \bea
  \rho_{dw} &=& 2 (b_{e}^{\dagger} b_{g}  +b_{g}^{\dagger} b_{e}   )
  \eea
  Therefore, the two checkerboard orders correspond to in-phase and out-of-phase superpositions of the $b_{g}$ mode and the $b_{e}$ mode.
   To capture the ordered states, we define the correlation function 
\bea
c_{dw}(t)&=& \frac{1}{N_{dw}}  \langle \rho_{dw}(t_{0}+t) \rho_{dw}(t_{0})\rangle_{t_{0}}\label{corrdw}
\eea

We evaluate the correlation functions $c_{ph}(t)$ and $c_{dw}(t)$ within the semiclassical approximation, i.e. we use
\bea
c_{ph}(t)&=& \frac{1}{\tilde{N}_{ph}} \langle \alpha^{*}(t_{0}+t) \alpha(t_{0})\rangle_{t_{0}}\label{corralpha}
\eea
and 
\bea
c_{dw}(t)&=& \frac{1}{\tilde{N}_{dw}}  \langle \tilde{\rho}_{dw}(t_{0}+t) \tilde{\rho}_{dw}(t_{0})\rangle_{t_{0}}\label{corrdw}
\eea
with
\bea
\tilde{\rho}_{dw} &=& 2 (\beta_{e}^{*} \beta_{g}  +\beta_{g}^{*} \beta_{e}   )
\eea
The presence or absence of an ordered or disordered  state corresponds to long-range or short-range order, respectively, of these correlation functions.

\subsection{Dicke phase transition}
 To ramp into a steady state above or below the critical pump intensity, we 
consider the following driving protocol $\epsilon_{p}(t)$, given by
\bea
\epsilon_{p}(t) &=& 
  \begin{cases}
      \epsilon_{f} (t/t_{f}) & \text{for\,\,} t<t_{f}\\
      \epsilon_{f} & \text{for\,\,} t\geq t_{f}
    \end{cases}\label{steadystate}
\eea
which is shown in Fig. \ref{trajsuprad} (a).
Throughout this review, we use parameters inspired by the experimental platform first described in Ref. \cite{Klinder2015}, and utilized in numerous later studies.  In this section, we use $\omega_{rec} = 2 \pi \times 3.55$ kHz, $\kappa = 2 \pi \times 4.5$ kHz, $\delta_{eff} = - 2 \pi \times 12.5$ kHz, $U_{0} = - 2 \pi \times 0.36$ Hz, and $N_{a} = 65\times 10^{3}$.  The protocol uses $t_{f} = 18$ ms, and two target values for $\epsilon_{f}$, specifically $\epsilon_{f} = 0.3$ below the critical pump intensity,  and $\epsilon_{f} = 3.3$ above the critical pump intensity.  
 In panel (b) of Fig. \ref{trajsuprad}, we show four randomly chosen trajectories, generated by the Heisenberg-Langevin equations given above. Specifically we show the real part of the photon amplitude $\alpha_{r} = \Re{\alpha}$, for the final pump intensity $\epsilon_{f} = 3.3$, i.e. larger than the critical value. Therefore, the trajectories either settle at a large positive or large negative value, corresponding to the spontaneous $\mathbb{Z}_{2}$ breaking.

   We evaluate the correlation function $c_{ph}(t)$ of the photon mode and $c_{dw}(t)$ of the density wave, for the steady state of the cavity-BEC system, after ramping up the driving intensity. 
    As indicated in Fig. \ref{trajsuprad} (b), we use the range of $[t_{min} = 40$ ms $, t_{max} = 45$ ms$]$ for averaging over $t_{0}$, which is in the steady state after the quench.
    
    In Fig. \ref{figsuprad} (a) we show the correlation function   of the photon mode $c_{ph}(t)$ for two final driving intensities, specifically for $\epsilon_{f}=3.3$ in blue, and for $\epsilon_{f} = 0.3$ in red.
    Similarly, In Fig. \ref{figsuprad} (b) we show the correlation function   of the density wave  $c_{dw}(t)$  for $\epsilon_{f}=3.3$ in blue, and  for $\epsilon_{f} = 0.3$ in red.
    For $\epsilon_{f}=3.3$, the steady state displays long-range order in both correlation functions. Both quantities are close to one for all times, each with an asymptotic value, i.e. condensate fraction, of around $\sim 1 - 10^{-4}$.
      For $\epsilon_{f}=0.3$, the steady state displays short-range order in both correlation functions, because the driving intensity is below the critical value. Both correlation functions decay with an exponential envelope, displaying oscillatory behavior. 

      \begin{figure}
\includegraphics[width=0.5\textwidth]{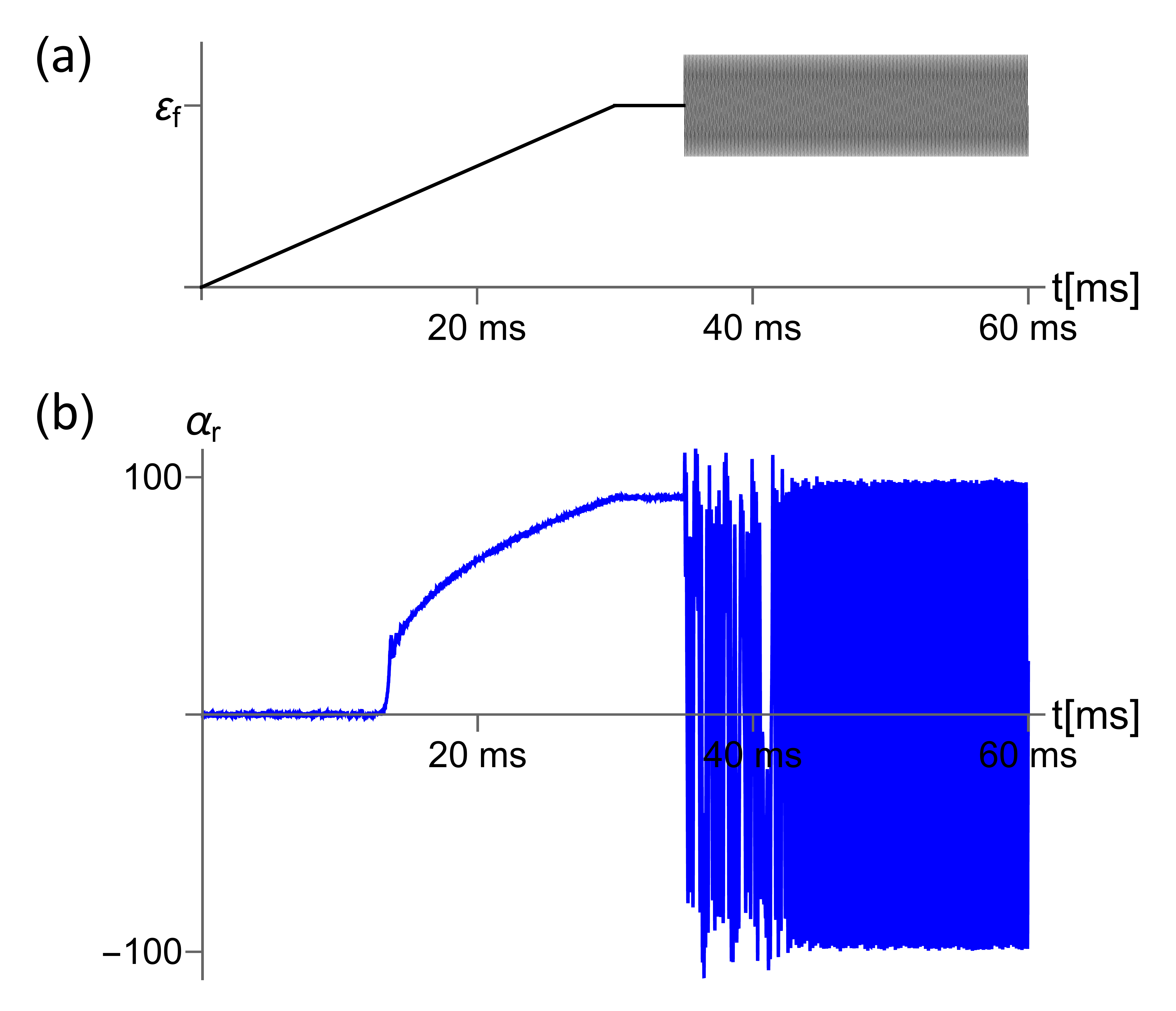}
\caption{In panel (a) we display the ramp and driving protocol. The intensity is first ramped up to a value $\epsilon_{f}=3.3$ in the superradiant state. After that, the intensity is modulated, with a relative amplitude $f_{0} =0.28$, and a driving frequency $\omega_{dr} =2\pi\times 10$ kHz. In panel (b) we display a trajectory of the real part of the photon field. After the intensity driving is turned on, the system first displays a transient regime, and then settles into a discrete time crystalline state, which is commensurate to the driving frequency. Specifically it is a period-doubling state.  \label{trajdtc}}
\end{figure} 

      \begin{figure}
\includegraphics[width=0.5\textwidth]{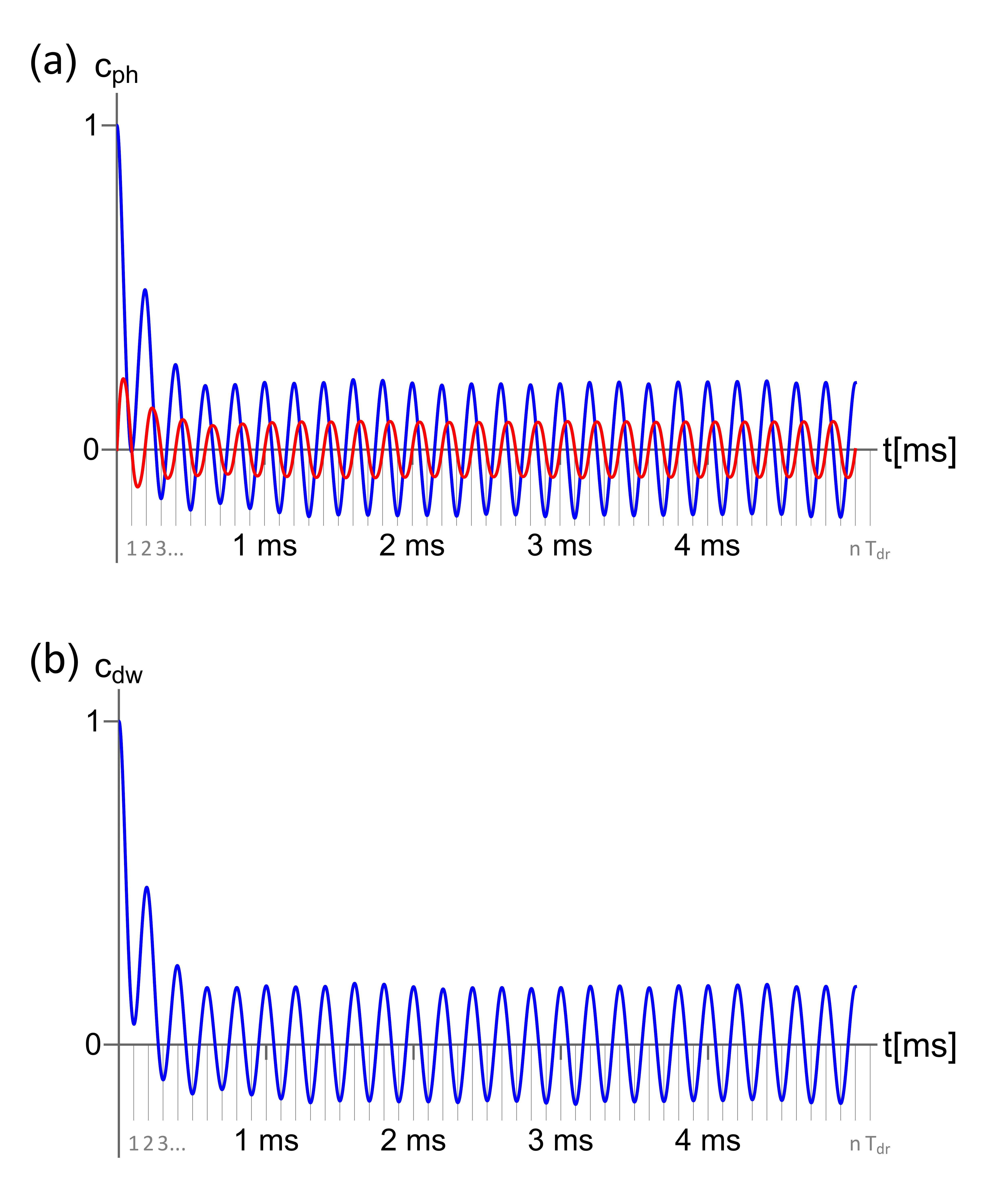}
\caption{In panel (a) we show the real (blue) and imaginary (red) part of the correlation function of the photon mode $c_{ph}(t)$ for the commensurate time crystalline state.  
 In panel (b) we show the correlation function of the checkerboard density order $c_{dw}(t)$ for the commensurate time crystal in blue.
  Additionally, we display $n T_{dr}$, i.e. integer multiples of the driving period $T_{dr}$ on the x-axis.
   Both correlation functions display oscillatory behavior, with an envelope that indicates long-range order. The comparison to the driving period shows that the oscillation is period-doubled compared to the driving period.
\label{figdtc}}
\end{figure} 

      \begin{figure}
\includegraphics[width=0.5\textwidth]{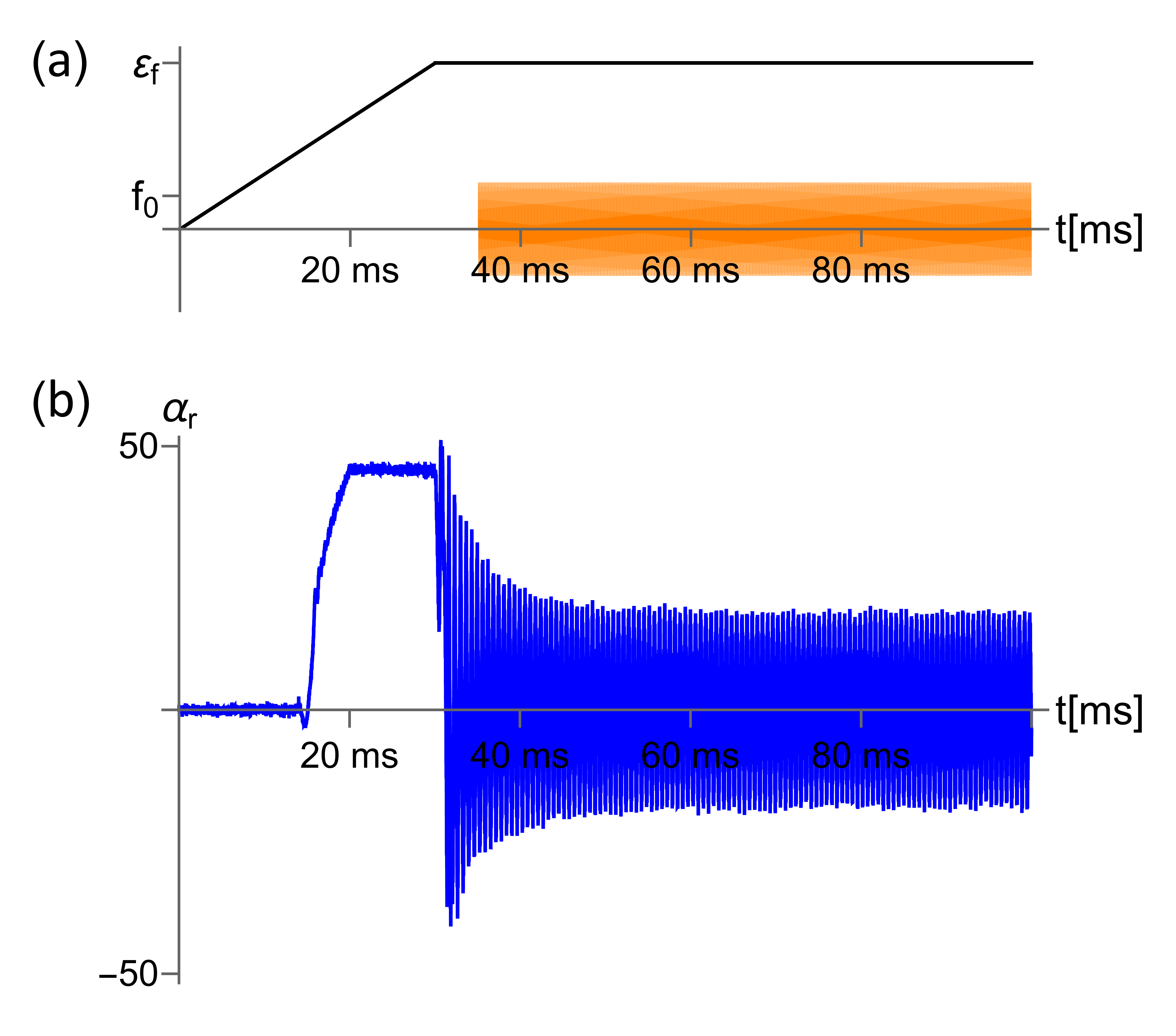}
\caption{In panel (a) we display the ramp and driving protocol. The intensity is first ramped up to a value $\epsilon_{f}=2.24$, within the superradiant state. After that, the phase is modulated with an amplitude $f_{0}=0.1$, depicted in yellow. In panel (b) we display a trajectory of the real part of the photon field. After the phase driving is turned on, the system displays a transient regime, and then settles into a time crystalline state, with an oscillatory motion that is incommensurate to the driving frequency.  \label{trajitc}}
\end{figure} 

       \begin{figure}
\includegraphics[width=0.5\textwidth]{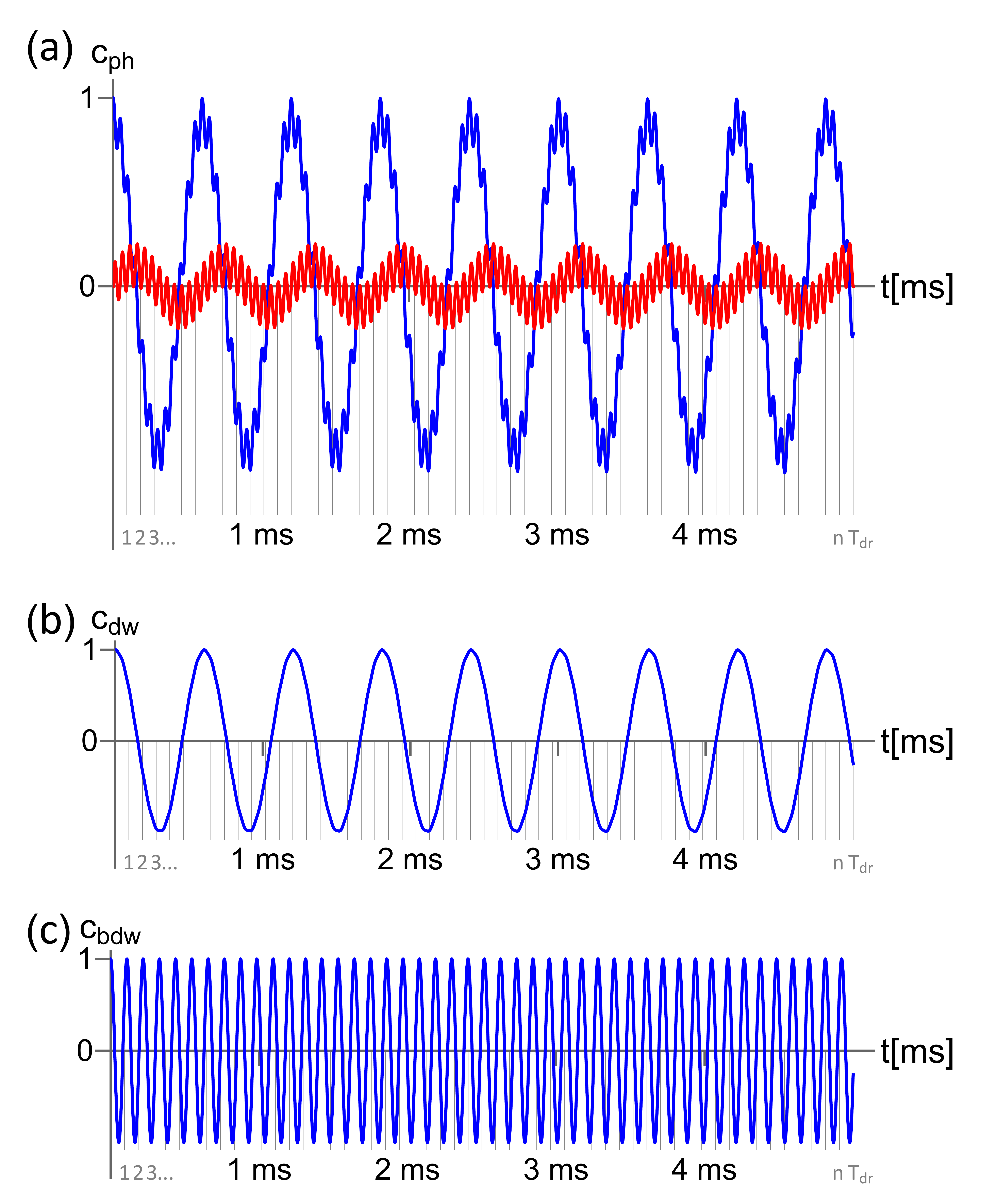}
\caption{In panel (a) we show the real (blue) and imaginary (red) part of the correlation function of the photon mode $c_{ph}(t)$ for the incommensurate time crystalline state. It displays an oscillatory behavior composed of two incommensurate frequencies. 
  In panel (b) we show the correlation function of the checkerboard density order $c_{dw}(t)$.
  In panel (c) we show the correlation function of the bond density order $c_{bdw}(t)$.
   The density correlation function displays primarily oscillations with the smaller frequency, the bond density correlation function displays oscillations with the larger frequency.
    All three correlation functions display  an envelope that indicates long-range order.
            Additionally, we display $n T_{dr}$, i.e. integer multiples of the driving period $T_{dr}$ on the x-axis.
                        The comparison to these multiples of $T_{dr}$ shows the incommensurability of the oscillation frequencies.
            \label{figitc}}
\end{figure} 
\section{Commensurate time crystal}
 We take the superradiant, ordered state as the starting point to drive the system into a 
 commensurate time crystal via  intensity modulation. 
 We implement this with
the following driving protocol $\epsilon_{p}(t)$, given by
\bea
\epsilon_{p}(t) &=& 
  \begin{cases}
      \epsilon_{f} (t/t_{f}) & \text{for\,\,} t<t_{f}\\
      \epsilon_{f} & \text{for\,\,} t_{dr}>t\geq t_{f}\\
       \epsilon_{f} [1+f_{0} \sin(\omega_{dr}(t-t_{dr}))] & \text{for\,\,} t \geq t_{dr}
    \end{cases}\label{intmod}
\eea
The protocol first follows the protocol given in Eq. \ref{steadystate}, to generate the steady state of the superradiant phase, with $t_{f} = 30$ ms and $\epsilon_{f} = 3.3$. 
 At time $t_{dr}$, the intensity modulation is performed, with a relative amplitude of $f_{0}$, and a driving frequency $\omega_{dr}$.
  
  In Fig. \ref{trajdtc} (a) we display the ramp and driving protocol, with $t_{dr} =35$ ms, $f_{0} = 0.28$, and $\omega_{dr} = 2\pi\times 10$ kHz.
   In Fig. \ref{trajdtc} (b), we show a trajectory of the real part of the cavity field  $\alpha_{r}$, to illustrate the dynamics.
      After the intensity driving is turned on, the system first displays transient dynamics, and relaxes into stable oscillatory motion, corresponding to a commensurate time crystal.
   We use $\delta_{eff} = 2\pi\times 18.5$ kHz, and $N_{a} = 65\times 10^{3}$, in this section, with all other parameters as stated before. 
      
      In Fig. \ref{figdtc} we display the correlation functions of the cavity mode and of the  density  wave order.
   In Fig. \ref{figdtc} (a) we show the real and imaginary part of $c_{ph}(t)$, and in Fig. \ref{figdtc} (b) we display $c_{dw}(t)$.
      To evaluate these correlation functions, we use $t_{min}= 50$ ms and $t_{max}=55$ ms.
    Both correlation functions show long-range order in an oscillatory fashion.
        In Fig. \ref{figdtc} (a) and (b) we indicate integer multiples of $n T_{dr}$ of the driving period $T_{dr} = 2\pi/\omega_{dr}$, on the time axis. 
     As is directly apparent, the oscillatory behavior of the correlation functions is twice the period of the driving protocol.
          Therefore it is a commensurate, period-doubling time crystal with $T_{TC} = 2 T_{dr}$.

\section{Incommensurate time crystal}
Next, we give an example for an incommensurate time crystal via phase modulation. 
As we see below, we will extend the two-model to a three-mode model in the atomic sector.
 We include the modulation of the phase of the pump laser by replacing $\cos(k y)$ with $\cos(k y + \phi(t))$. With this, $H_{A}$ is of the form 
 \bea
 \frac{H_{A}}{\hbar}&=& \int d^{2}r \Psi^{\dagger}(\br) \Big(  -\frac{\hbar \nabla^{2}}{2 m}\nonumber\\
&& - \omega_{rec} \epsilon_{p} \cos^{2}(k y + \phi(t)) \Big) \Psi(\br)
\eea
 and $H_{AC}$ is of the form
\bea
 && \frac{H_{AC}}{\hbar}= \int d^{2}r \Psi^{\dagger}(\br) \Big( U_{0} a^{\dagger} a \cos^{2}(k z)\nonumber\\
&&  +  2 g \sqrt{\epsilon_{p}} \cos(k y + \phi(t))\cos(k z)(a^{\dagger} + a)  \Big) \Psi(\br)
\eea
In momentum space representation, we obtain
 \bea
\!\!\!&&\frac{H_{A}}{\hbar}= \sum_{\bp} \Big( \omega_{p}  - \frac{\omega_{rec} \epsilon_{p}}{2} \Big) b_{\bp}^{\dagger} b_{\bp}\nonumber\\
\!\!\!&& - \frac{\omega_{rec} \epsilon_{p}}{4}  \sum_{\bp} ( e^{-2i\phi(t)} b_{\bp}^{\dagger} b_{\bp+2k\be_{y}}
 +  e^{2i\phi(t)} b_{\bp}^{\dagger} b_{\bp-2k\be_{y}})
\eea
and
\bea
 &&\frac{H_{AC}}{\hbar} = \frac{U_{0}}{2} a^{\dagger} a \sum_{\bp} b_{\bp}^{\dagger} b_{\bp}
 + \frac{U_{0}}{4} a^{\dagger} a \sum_{\bp} b_{\bp}^{\dagger} (b_{\bp +2 k \be_{z}} + b_{\bp - 2 k \be_{z}})\nonumber\\
&& + \frac{\text{sgn}(U_{0}) \sqrt{\omega_{rec} |U_{0}| \epsilon_{p}}}{4} (a^{\dagger} + a)
 \sum_{\bp} b_{\bp}^{\dagger} ( e^{-i\phi(t)} (b_{\bp + k\be_{y} + k \be_{z}}\nonumber\\
&& + b_{\bp + k\be_{y} - k \be_{z}}) + e^{i\phi(t)} (b_{\bp - k\be_{y} + k \be_{z}}
 + b_{\bp - k\be_{y} - k \be_{z}}))
\eea
We note that the term proportional to $a^{\dagger} + a$ couples the condensate mode $b_{\bp=0}$ differently to the modes  $b_{k\be_{y} \pm k \be_{z}}$, than to the modes $b_{-k\be_{y} \pm k \be_{z}}$, due to the phase factors $\exp(\pm i \phi(t))$. 
 This suggests that the phase modulation activates an additional set of momentum modes, beyond $b_{g}$ and $b_{e}$.

\subsection{Three-mode expansion}
We derive an effective three-mode model based on this model, by including 
 \bea
 b_{s_{1}}&=& 
 \frac{1}{2}( b_{k\be_{y} + k \be_{z}}
  + b_{ k\be_{y} - k \be_{z}}\nonumber\\
  && - b_{ -k\be_{y} + k \be_{z}} - b_{ - k\be_{y} - k \be_{z}} )
 \eea
in addition to $a$, $b_{g}$, and $b_{e}$, as in Eq. \ref{H2m}. 
With this additional mode, the Hamiltonian takes the form
   \bea
&& \frac{H_{3m,1}}{\hbar} =  -\delta_{eff} a^{\dagger} a + 2 \omega_{rec} ( b_{e}^{\dagger} b_{e} + b_{s_{1}}^{\dagger} b_{s_{1}})\nonumber\\
&&  + \frac{U_{0}}{4} a^{\dagger} a (b_{e}^{\dagger} b_{e} +b_{s_{1}}^{\dagger} b_{s_{1}})\nonumber\\
&& - \frac{\omega_{rec} \epsilon_{p}}{4} \Big( \cos(2\phi(t)) (b_{e}^{\dagger} b_{e} - b_{s_{1}}^{\dagger} b_{s_{1}})\nonumber\\
 &&+ i \sin(2\phi(t)) (b_{s_{1}}^{\dagger} b_{e} - b_{e}^{\dagger} b_{s_{1}}) \Big)\nonumber\\
 && + g \sqrt{\epsilon_{p}} (a^{\dagger} + a) \Big( \cos(\phi(t)) (b_{g}^{\dagger} b_{e} + b_{e}^{\dagger} b_{g} )\nonumber\\
 && + i \sin(\phi(t)) (b_{s_{1}}^{\dagger} b_{g} - b_{g}^{\dagger} b_{s_{1}} )\Big)
 \eea
 We note that for $\phi(t)=0$, the Hamiltonian reduces to the Hamiltonian given in Eq. \ref{H2m}, and the state $b_{s_{1}}$ decouples. 
  We use the notation $H_{3m,1}$ for this model to distinguish it from the model $H_{3m,2}$ that we use in the context of the continuous time crystal below. 
  $H_{3m,2}$ is based on three modes in the atomic sector as well, but uses a mode $b_{s_{2}}$ that is different than  the mode $b_{s_{1}}$ that we use here.

 \subsection{Semiclassical dynamics}
   The Heisenberg-Langevin equations are
\bea
\frac{d \alpha}{d t} &=&  i \delta_{eff} \alpha - \kappa \alpha -i \frac{U_{0}}{4} (\beta_{e}^{*}\beta_{e} + \beta_{s_{1}}^{*}\beta_{s_{1}} )\alpha\nonumber\\ 
&& - i g \sqrt{\epsilon_{p}}
 \Big( \cos(\phi(t)) (\beta_{g}^{*} \beta_{e} + \beta_{e}^{*} \beta_{g} )\nonumber\\
 && + i \sin(\phi(t)) (\beta_{s_{1}}^{*} \beta_{g} - \beta_{g}^{*} \beta_{s_{1}} )\Big) + \xi\\
\frac{d \beta_{g}}{d t} &=& - i g \sqrt{\epsilon_{p}} (\alpha^{*} + \alpha) (\cos(\phi(t))\beta_{e}\nonumber\\
&& - i \sin(\phi(t)) \beta_{s_{1}})\\
\frac{d \beta_{e}}{d t} &=& - 2 i \omega_{rec} \beta_{e} -i \frac{U_{0}}{4} \alpha^{*} \alpha \beta_{e}\nonumber\\ 
&& + i \frac{\omega_{rec} \epsilon_{p}}{4} \Big(\cos(2\phi(t)) \beta_{e} - i \sin(2\phi(t)) \beta_{s_{1}} \Big)\nonumber\\
&& - i g \sqrt{\epsilon_{p}} (\alpha^{*} + \alpha) \cos(\phi(t)) \beta_{g}\\
\frac{d\beta_{s_{1}}}{d t} &=& - 2 i \omega_{rec} \beta_{s_{1}} - i \frac{U_{0}}{4} \alpha^{*} \alpha \beta_{s_{1}}\nonumber\\
&& - i \frac{\omega_{rec} \epsilon_{p}}{4} \Big( \cos(2\phi(t)) \beta_{s_{1}} - i \sin(2\phi(t)) \beta_{e}\Big)\nonumber\\
&& + g \sqrt{\epsilon_{p}} (\alpha^{*} + \alpha) \sin(\phi(t)) \beta_{g}
\eea
As before, and throughout this review, the cavity mode dynamics entails photon decay via the the term $- \kappa\alpha$, and decay backaction via the noise term $\xi$.

\subsection{Correlation functions}
To capture the density dynamics  of the time crystalline state, we define the bond density wave operator:
\bea
\rho_{bdw} &=& i (\rho_{k \be_{y} +k \be_{z}} + \rho_{k \be_{y} -k \be_{z}}\\
&&- \rho_{-k \be_{y} +k \be_{z}} - \rho_{-k \be_{y} -k \be_{z}})
\eea
 with $\rho_{k} = \sum_{p} b_{p}^{\dagger} b_{p+k}$, as above, where the prefactor $i$ is introduced for the operator to be Hermitian. 
  The expectation value of this density mode takes maximal values for checkerboard patterns as well. However, in contrast to 
 the checkerboard mode $\rho_{dw}$, the mode  $\rho_{bdw}$ describes a bond-ordered density pattern, with maxima on the lattice bonds along the pump direction $y$. 
  Written in the three-mode expansion, we have
  \bea
  \rho_{bdw} &=& 2 i (b_{g}^{\dagger} b_{s_{1}}  - b_{s_{1}}^{\dagger} b_{g}   )
  \eea
We define the correlation function
\bea
c_{bdw}(t)&=& \frac{1}{N_{dbw}}  \langle \rho_{bdw}(t_{0}+t) \rho_{bdw}(t_{0})\rangle_{t_{0}}\label{corrbdw}
\eea
analogously to $c_{dw}(t)$ and $c_{ph}(t)$.
 As for these correlation functions, we evaluate $c_{bdw}(t)$ in the semiclassical approximation:
\bea
c_{bdw}(t)&=& \frac{1}{\tilde{N}_{bdw}}  \langle \tilde{\rho}_{bdw}(t_{0}+t) \tilde{\rho}_{bdw}(t_{0})\rangle_{t_{0}}\label{corrbdw}
\eea
with
\bea
\tilde{\rho}_{bdw} &=& 2 i (\beta_{g}^{\dagger} \beta_{s_{1}}  - \beta_{s_{1}}^{\dagger} \beta_{g}   )
\eea
 The resulting incommensurate time crystal displays long-range order in the three correlation functions $c_{ph}(t)$, $c_{dw}(t)$,  and $c_{bdw}(t)$, as we see in the example below. 

      \begin{figure}
\includegraphics[width=0.5\textwidth]{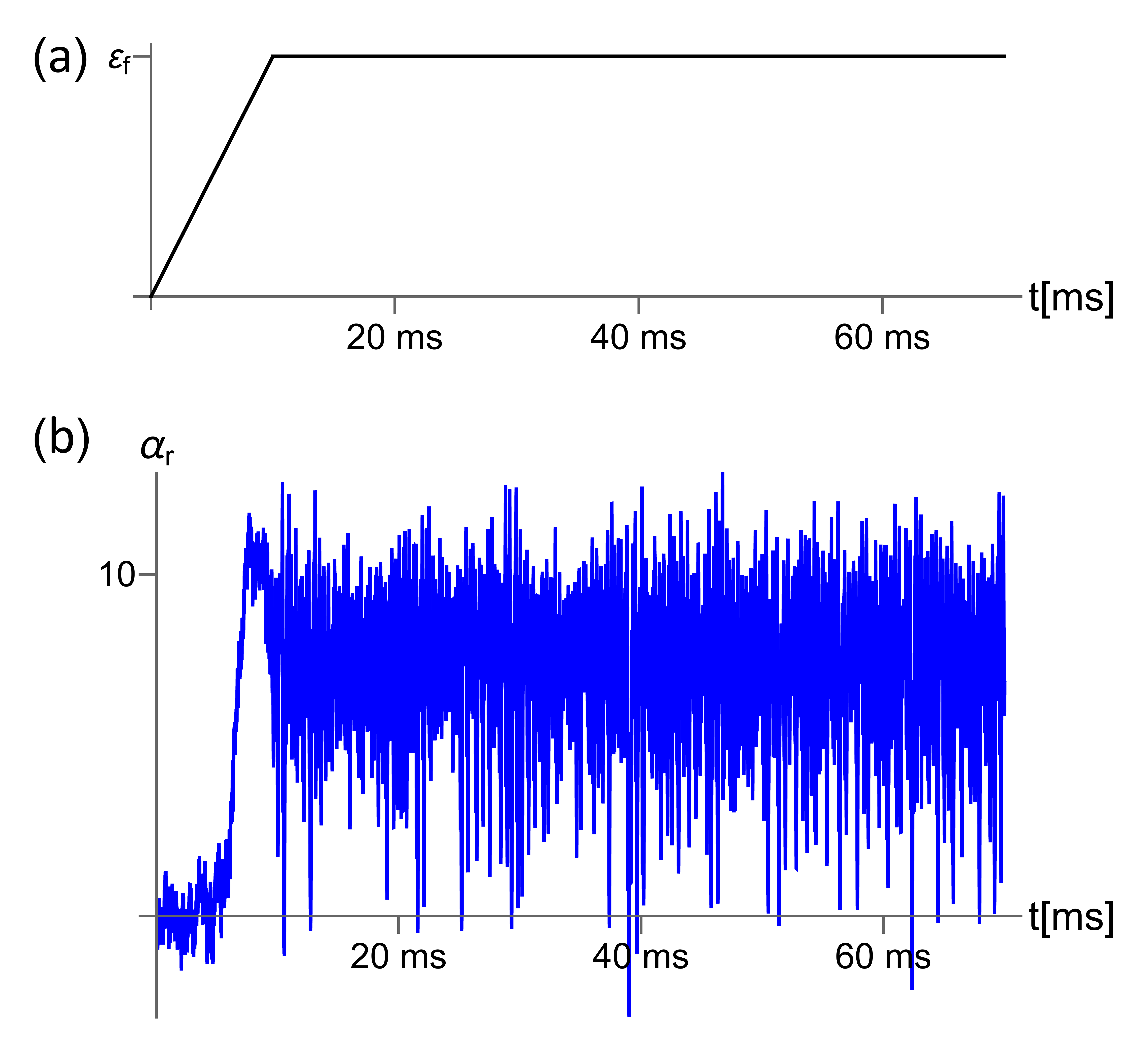}
\caption{
 In panel (a) we show the ramp into the continuous time crystalline state. It is qualitatively the same as in Fig. \ref{trajsuprad} (a), However, we use a different set of system parameters, in particular smaller detuning.
 In panel (b) we show a single trajectory $\alpha_{r}(t)$. In contrast to the trajectories shown in Fig. \ref{trajsuprad} (b), the trajectory oscillates and fluctuates. Evaluating the correlation functions of this state, shows that this is a continuous time crystal. 
 \label{trajctc}}
\end{figure}

      \begin{figure}
\includegraphics[width=0.5\textwidth]{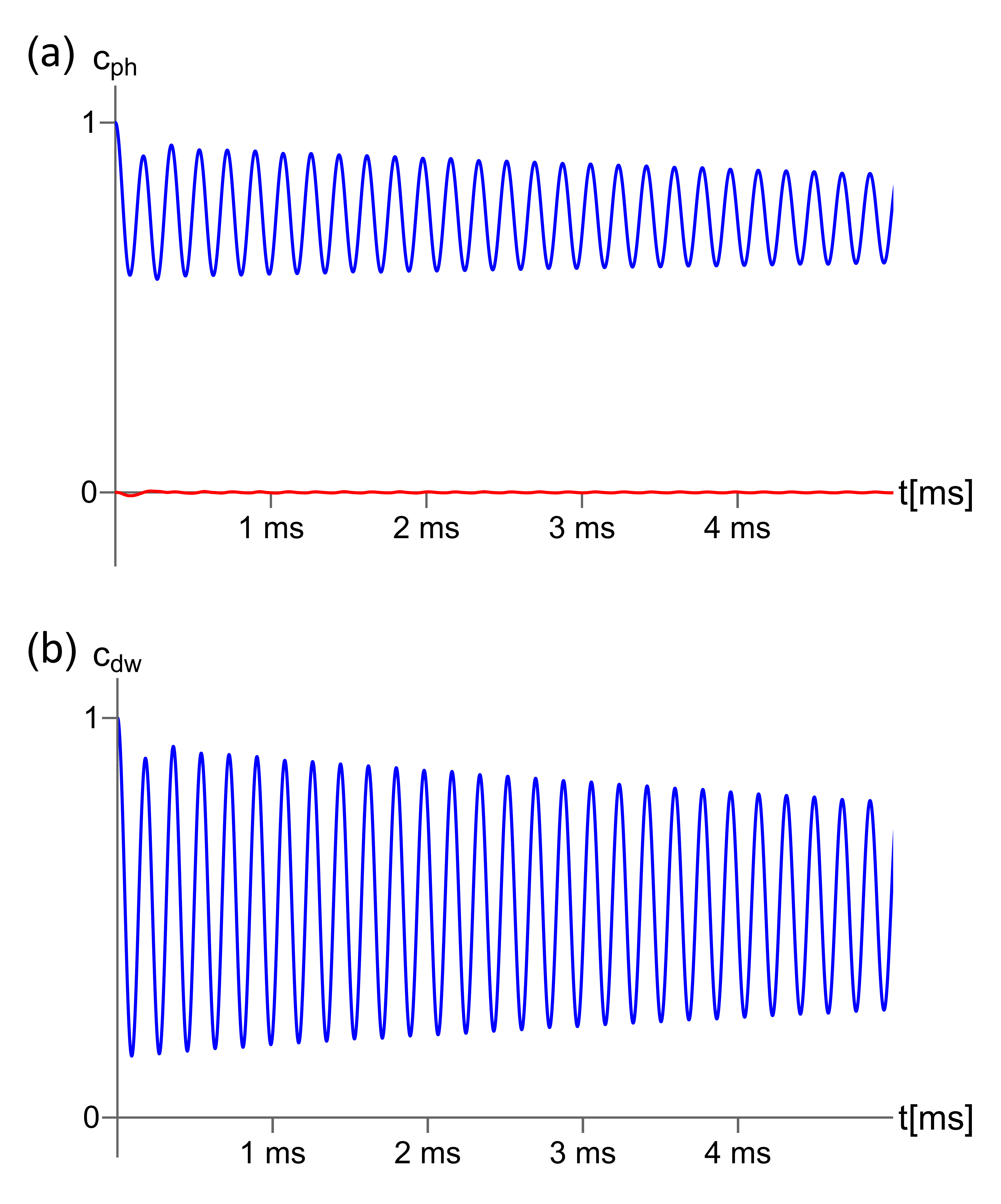}
\caption{
 In panel (a) we show the real and imaginary part of $c_{ph}(t)$. Both show oscillatory motion, with a slowly declining envelope. 
The amplitude of the imaginary part is much smaller than the real part, but does not average to zero.
 In panel (b) we show $c_{dw}(t)$, which also shows oscillatory motion with a slowly decaying envelope. In the thermodynamic limit, the continuous time crystalline state develops long-range order. 
\label{figctc}}
\end{figure} 

\subsection{Phase transition}
To drive  the system into the incommensurate  time crystalline state, we first ramp into the superradiant state, following the protocol given by  Eq. \ref{steadystate}.
 After that, we modulate the phase $\phi(t)$ of the pump laser via the protocol: 
\bea
\phi(t) &=& 
  \begin{cases}
      0 & \text{for\,\,} t<t_{dr}\\
       f_{0} \sin(\omega_{d}(t-t_{dr})) & \text{for\,\,} t \geq t_{dr}
    \end{cases}\label{phasemod}
\eea
 The phase modulation is turned on at time $t_{dr}$, with a frequency of $\omega_{dr}$, and an amplitude $f_{0}$.
  In Fig. \ref{trajitc} (a) we show the driving protocol. 
   In black, we show the pump intensity, with $t_{f} = 20$ ms, and $\epsilon_{f} = 2.24$.
    In orange, we show the phase protocol $\phi(t)$ of Eq.  \ref{phasemod}. 
     At $t_{dr} =30$ ms, we turn on the phase modulation, with $\omega_{dr} = 2 \pi\times 10.8$ kHz, and $f_{0}=0.1$.
      We use $\kappa = 2\pi\times 3.6$ kHz, $\delta_{eff} = -2\pi\times 22$ kHz, and $N_{a} = 60\times 10^{3}$, and all other parameters as above. 
       In Fig. \ref{trajitc} (b) we show a trajectory $\alpha_{r}(t)$.
      The system first develops a superradiant state, due to the intensity ramp-up above the critical pump intensity.
      After the phase modulation is turned on at time $t_{dr}$,  the system first displays transient dynamics, and then settles into a time crystalline state.
  
   In Fig. \ref{figitc} (a) we show the real and imaginary part of $c_{ph}(t)$, in blue and orange, respectively.
   The integer multiples $n T_{dr}$ of the driving period are indicated on the time axis.
  The time dependence of $c_{ph}(t)$ is dominated by two frequencies, both of which are incommensurate with $T_{dr}$.
  In particular, one frequency is significantly lower than $\omega_{dr}$, corresponding to a period $T_{TC}$ much larger than $T_{dr}$.
  This long period $T_{TC}$ is also visible in $c_{dw}(t)$, shown in Fig. \ref{figitc} (b). The periodic change of the real part of $c_{ph}(t)$ and of $c_{dw}(t)$ implies that the system oscillates between the two superradiant states, but not with a period-doubled frequency, but a low, incommensurate frequency, thus forming an incommensurate time crystal.
  The correlation function $c_{bdw}(t)$, shown in  Fig. \ref{figitc} (c), oscillates at the higher frequency, which is close to the driving frequency.
  The phase modulation drives the checkerboard ordered state along the bonds along the pump direction, resulting in this signature in $c_{bdw}(t)$. 
   The three correlation functions $c_{ph}(t)$, $c_{dw}(t)$, and $c_{bdw}(t)$, are evaluated based on the time interval of $t_{min}=90$ ms and $t_{max}=95$ ms.
  The nonlinearities of the system convert the driving frequency  into the  oscillation frequencies, via a parametric mechanism \cite{Skulte2021}.

\section{Continuous time crystal}\label{ctcsect}
 We now extend the continuously driven two-mode model, given by Eq. \ref{H2m},  to a three-mode model, which differs from the one introduced for the incommensurate time crystal, by including the state
 \bea
 b_{s_{2}}&=& \frac{1}{\sqrt{2}}(b_{2 k \be_{z}} +b_{-2 k \be_{z}} )
 \eea
We do not consider additional driving, such as intensity and phase driving, see Fig. \ref{ovtc}. So the ramp-up protocol is of the form given by Eq. \ref{steadystate}. 
  With the additional state $b_{s_{2}}$, the Hamiltonian takes the form
   \bea
 \frac{H_{3m,2}}{\hbar} &=&  \frac{H_{2m}}{\hbar}  + 4 \omega_{rec} b_{s_{2}}^{\dagger} b_{s_{2}}\nonumber
  + \frac{U_{0}}{\sqrt{2}} a^{\dagger} a (b_{g}^{\dagger} b_{s_{2}}  + b_{s_{2}}^{\dagger} b_{g})\nonumber\\
&& + \frac{g}{\sqrt{2}} \sqrt{\epsilon_{p}} (a^{\dagger} + a) (b^{\dagger}_{s_{2}} b_{e} + b_{e}^{\dagger} b_{s_{2}} )
 \eea
 The equations of motion take the form 
\bea
\frac{d \alpha}{d t} &=&  i \delta_{eff} \alpha - \kappa \alpha - i g \sqrt{\epsilon_{p}} (\beta_{g}^{*} \beta_{e} + \beta_{e}^{*} \beta_{g})\nonumber\\
&& -i \frac{U_{0}}{4} \beta_{e}^{*} \beta_{e} \alpha 
 - i \frac{U_{0}}{\sqrt{2}} (\beta_{g}^{*} \beta_{s_{2}}  + \beta_{s_{2}}^{*} \beta_{g}) \alpha\nonumber\\
 && - i \frac{g \sqrt{\epsilon_{p}}}{\sqrt{2}}   (\beta^{*}_{s_{2}} \beta_{e} + \beta_{e}^{*} \beta_{s_{2}} )
+ \xi\label{ctcalpha}\\
\frac{d \beta_{g}}{d t} &=& - i g \sqrt{\epsilon_{p}} (\alpha^{*} + \alpha) \beta_{e}
 - i \frac{U_{0}}{ \sqrt{2}} \alpha^{*} \alpha \beta_{s_{2}}\label{ctcbetag}\\
\frac{d \beta_{e}}{d t} &=& - 2 i \omega_{B} \beta_{e} - i g \sqrt{\epsilon_{p}} (\alpha^{*} + \alpha) \beta_{g}\nonumber\\
&& -i \frac{U_{0}}{4} \alpha^{*} \alpha \beta_{e} - i \frac{g \sqrt{\epsilon_{p}}}{\sqrt{2}} (\alpha^{*} + \alpha) \beta_{s_{2}}\label{ctcbetae}\\
\frac{d \beta_{s_{2}}}{d t} &=& - 4 i \omega_{rec} \beta_{s_{2}} - i \frac{U_{0}}{\sqrt{2}} \alpha^{*}\alpha \beta_{g}\nonumber\\ 
&& - i \frac{g}{\sqrt{2}} \sqrt{\epsilon_{p}} (\alpha^{*} + \alpha) \beta_{e}\label{ctcbetas2} 
\eea
With the third mode $b_{s_{2}}$, the density operator takes the form
  \bea
  \rho_{dw} &=& 2 (b_{e}^{\dagger} b_{g}  +b_{g}^{\dagger} b_{e}   ) + \frac{1}{\sqrt{2}}(b_{e}^{\dagger} b_{s_{2}} +b_{s_{2}}^{\dagger} b_{e}  )
  \eea
   We use this representation to evaluate $c_{dw}(t)$, in the semiclassical approximation.
    
 In Fig. \ref{trajctc} (a)   we show the pump protocol for $t_{f}= 10$ ms and $\epsilon_{f} = 2.0$. 
  In Fig. \ref{trajctc} (b) we show a trajectory $\alpha_{r}(t)$.  As parameters we use $\kappa = 2\pi\times 3.6$ kHz, $\delta_{eff} = -2\pi\times 1.5$ kHz,and $N_{a} = 40\times 10^{3}$.
   We observe that the system does not form a stable superradiant state, but rather a continuous time crystalline state.
       We note that the detuning $\delta_{eff}$ is smaller than in the example shown in Fig. \ref{trajsuprad}, which is one of the criteria for generating this state.
   Additionally, we either have to include the mode $b_{s_{2}}$, or integrate out $b_{s_{2}}$, and include the resulting nonlinear coupling, cp. Ref. \cite{Skulte2024}. 
   
   In Fig. \ref{figctc}  we show $c_{ph}(t)$ and $c_{dw}(t)$, evaluated for $t_{min}= 40$ ms, and $t_{max} = 55$ ms.
   
   We observe a slowly decaying oscillatory state, which achieves long-range order in the thermodynamic limit, as we discuss below.
    The magnitude of real-part of $c_{ph}(t)$ and of $c_{dw}(t)$ oscillate around a symmetry-broken state. So, in contrast to the previous two time crystalline states, the system does not oscillate between two symmetry-broken states.

\subsection{Thermodynamic limit}
In this section we discuss the thermodynamic limit of cavity-BEC systems, represented in a few-mode model. In this limit, the continuous time crystal described in the previous section, and shown in Fig. \ref{figctc}, displays long range order.
 For this purpose, we utilize the quantum-classical rescaling discussed in \cite{nowoczyn2025universalquantummeltingquasiperiodic}.
 We consider the system of equations \ref{ctcalpha} - \ref{ctcbetas2}, and generate the set of dynamical variables
\bea
\alpha' &=& \sqrt{\aleph} \alpha\\
\beta'_{g, e, s_{2}} &=& \sqrt{\aleph} \beta_{g, e, s_{2}}
\eea
and  rescale the parameters of the equations of motion via
\bea
g'&=& \frac{g}{\sqrt{\aleph}}\\
U_{0}' &=& \frac{U_{0}}{\aleph}\\
\kappa' &=& \frac{\kappa}{\aleph}
\eea
The dynamical variables $\alpha'$ and $\beta'_{g, e, s_{2}}$ solve the set of equations with the parameters $g'$, $U_{0}'$, and $\kappa'$.
 We note that the back action noise that influences the dynamics of $\alpha'$ has the same effective decay rate $\kappa$, due to the cancelation of the rescaling prefactors for this quantity. The system dynamics of $\alpha'$ and $\beta'_{g, e, s_{2}}$ therefore correspond to a cavity with the same characteristics.
  
\begin{figure}
\includegraphics[width=0.5\textwidth]{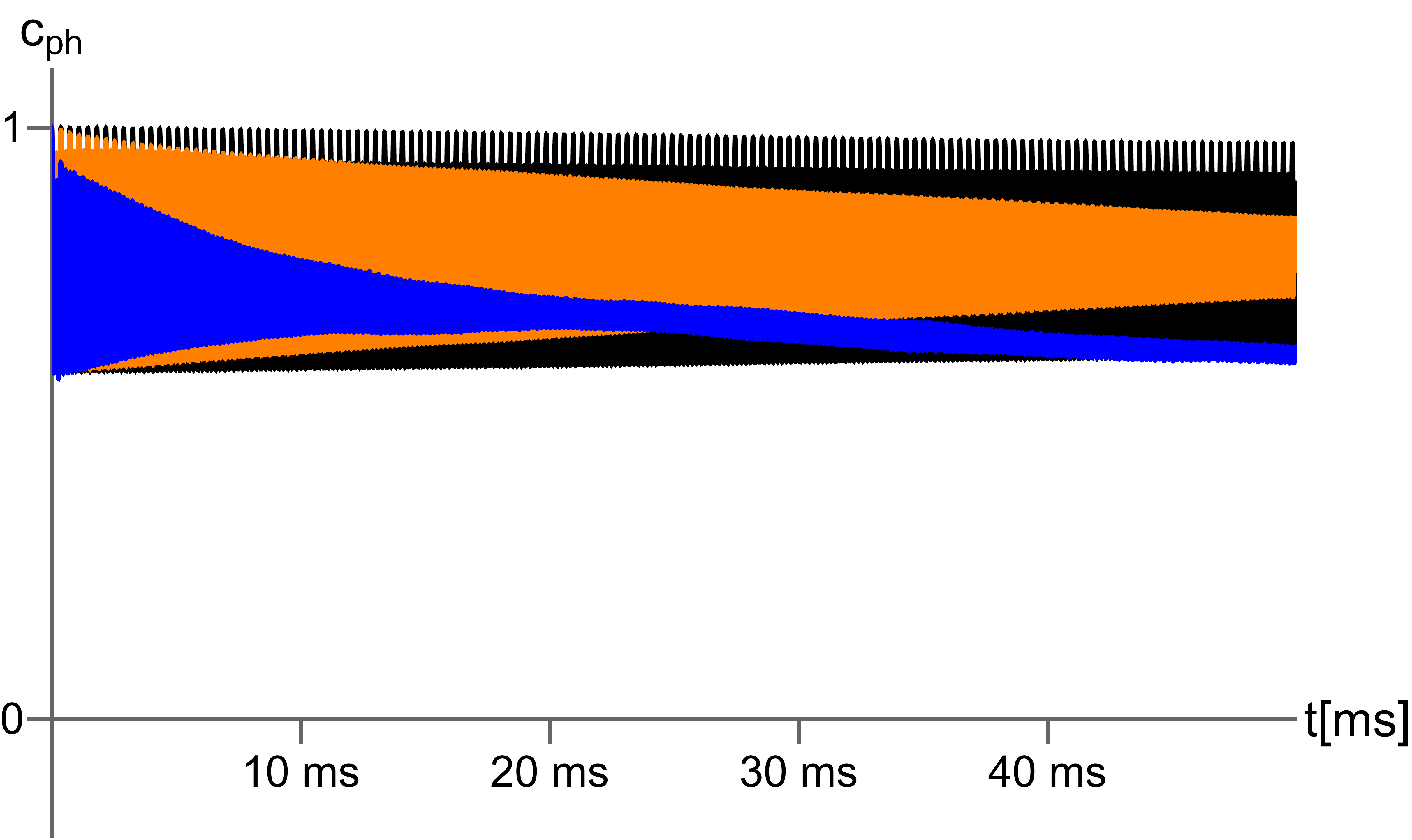}
\caption{
  We show the real part of the correlation function $c_{ph}(t)$, as the system approaches the thermodynamic limit.
  In blue, we display the same result as in Fig. \ref{figctc}, for a longer time interval.
   In orange, we use $\aleph = 10$, and  in black, we use $\aleph = 100$.
   The correlation function approaches long-range order for increasing $\aleph$, or increasing atom number. 
\label{ctcthermo}}
\end{figure}

  The rescaling results in a  rescaling of the total atom number $N_{a}$ via $N_{a}' \approx \aleph N_{a}$, due to the rescaling $\beta'_{g, e, s_{2}} = \sqrt{\aleph} \beta_{g, e, s_{2}}$.
   We note that the quantum-classical rescaling is set up such that it leaves the classical dynamical regime of the equations of motion invariant. In particular, the limit cycle dynamics of the fluctuation-free limit of the dynamics, i.e. ignoring the noise contribution, is unchanged.
      Therefore, the quantum-classical rescaling generates a large-atom limit that leaves the qualitative dynamical regime unchanged.
      In Fig.  \ref{ctcthermo} we display the real-part of the correlation function  $c_{ph}(t)$ for $\aleph = 1$, $\aleph = 10$, and $\aleph = 100$, which corresponds to $N_{a} = 4\times 10^{4}$, $N_{a} = 4\times 10^{5}$, and $N_{a} = 4\times 10^{6}$, respectively.
   We observe a stabilization of the correlation function, indicating that long-range order is achieved in the thermodynamic limit.

      \begin{figure}
\includegraphics[width=0.42\textwidth]{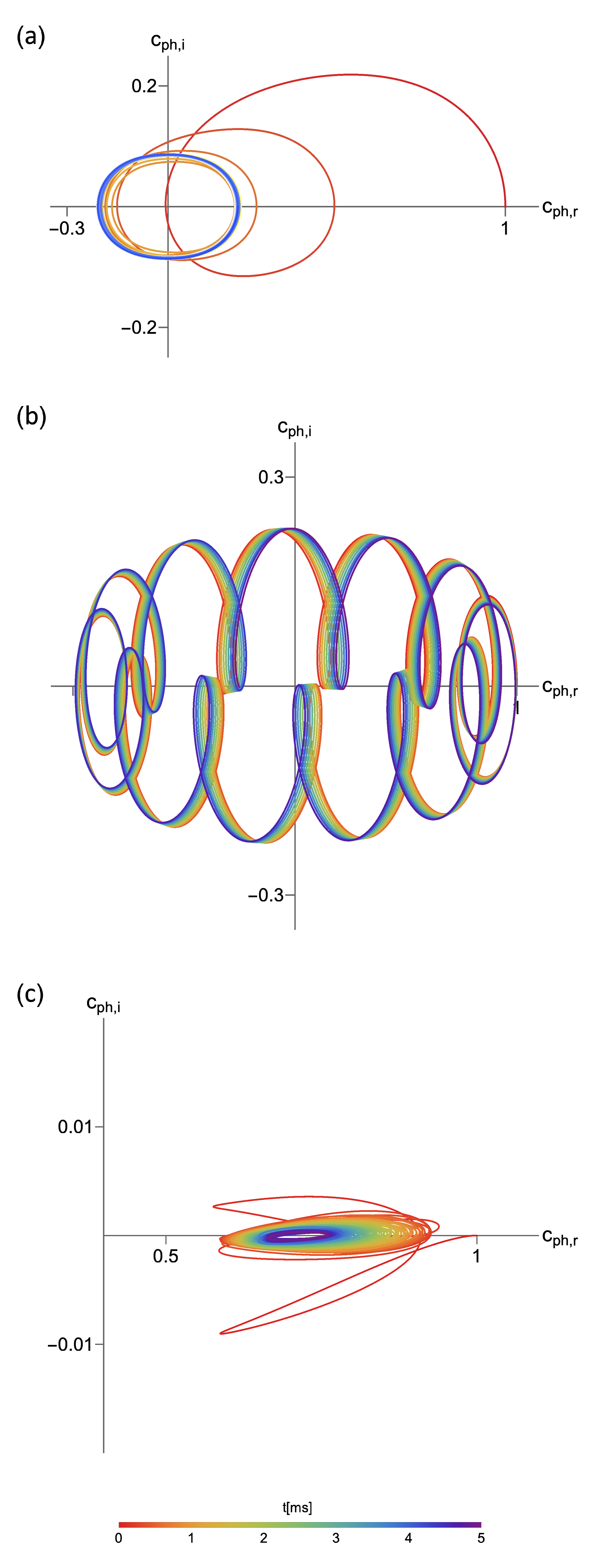}
\caption{
We show a parametric representation of $c_{ph}(t)$ for the three examples for time crystalline states.
 In panel (a) we show the commensurate time crystal, corresponding to $c_{ph}(t)$ shown in Fig. \ref{figdtc} (a).
 In panel (b) we show the incommensurate time crystal, corresponding to $c_{ph}(t)$ shown in Fig. \ref{figitc} (a). 
   In panel (c) we show the continuous time crystal, corresponding to $c_{ph}(t)$ shown in Fig. \ref{figctc} (a).
    In all cases, we show the first 5 ms of the correlation function, as indicated in the color bar. 
 \label{para}}
\end{figure} 
\section{Conclusions}
In conclusion, we have presented three examples for time crystalline states in a cavity-BEC system.
 For each, we have derived a minimal, few-mode model.
  We have presented the correlation function of the cavity mode and of the relevant density components, for each of the examples.
   The time-crystalline states manifest themselves as long-range order features in these correlation functions.
   
   Pictorially, the three time crystalline states are depicted in Fig. \ref{ovtc}.
  The first example is a commensurate time crystal, induced by intensity modulation. The minimal model that we employ, is a model composed of the cavity mode and two atomic modes.
   The correlation function of the cavity mode in the commensurate time crystalline state is shown in Fig. \ref{figdtc} (a).
    The real and imaginary part of the correlation function display oscillations, with an envelope that is consistent with long-range order.
     The oscillation period of the correlation function is twice the driving frequency. Therefore the time crystalline state is a period-doubling state.
    This period-doubling property is visible in the correlation function of the density-wave order, i.e. the checkerboard order, in  Fig. \ref{figdtc} (b).
     It switches sign for each driving period, consistent with period-doubling oscillations.
     
     The second example is an incommensurate time crystal, induced by phase modulation. 
     We employ a minimal model composed of the cavity mode and three atomic modes.
         The correlation function of the cavity mode in the incommensurate time crystalline state is shown in Fig. \ref{figitc} (a).
      The real and imaginary part of the correlation function display oscillations that are composed of two frequencies, which are both incommensurate to the driving frequency.
       In particular, there is a low-frequency motion in the real-part of the cavity  correlation function $c_{ph}(t)$, as well as in the density correlation function $c_{dw}(t)$, that shows oscillations between the two superradiant states, at an incommensurate frequency.
        
        The third example is a continuous time crystal. For this time crystal, there is no periodic driving in the rotating frame, but only constant driving.
        For sufficiently small detuning, and for intermediate pump intensity, the system oscillates near a superradiant state, rather forming a stable superradiant state.
         This constitutes a continuous time crystal, in a driven-dissipative system.
         
         In Fig. \ref{para}, we summarize the cavity correlation functions $c_{ph}(t)$ for the three time crystalline states, now shown as a parametric representation in the complex plane.
          The commensurate time crystal shows stable oscillations around the origin, see Fig. \ref{para} (a). The incommensurate time crystal shows a trajectory around the origin, composed of two frequencies, see Fig. \ref{para} (b).
          The continuous time crystal has a very small imaginary component, and shows an oscillation near the real axis.
           The continuous time crystalline motion of a system for a large atom number, with $\aleph=100$, is shown in Fig. \ref{paralarge}, as a comparison. 
          
          With the creation of the three time crystalline states presented in this review, and their experimental realization, we have demonstrated the suitability and versatility of atom-cavity systems for the realization, characterization and understanding of dynamical phenomena.  
      Going forward, the framework of fluctuating few-mode models for the minimal description of dynamical states, provides guidance for other environments and phenomena.     

\section{Acknowledgments}
 This work was funded by the Deutsche Forschungsgemeinschaft (DFG, German Research Foundation) ``SFB-925" project 170620586, the Cluster of Excellence ``Advanced Imaging of Matter" (EXC 2056), Project No. 390715994, ERDF of the European
Union and by ''Fonds of the Hamburg Ministry of Science,
Research, Equalities and Districts (BWFGB)'', and the Department of Science and Technology (DOST) as monitored by the Philippine Council for Industry, Energy, and Emerging Technology Research and Development (DOST-PCIEERD)
through Project No. 1214356.
 We thank Caroline Nowoczyn for valuable discussions.

\section{Appendix}

\subsection{Large atom limit}
\begin{figure}
\includegraphics[width=0.45\textwidth]{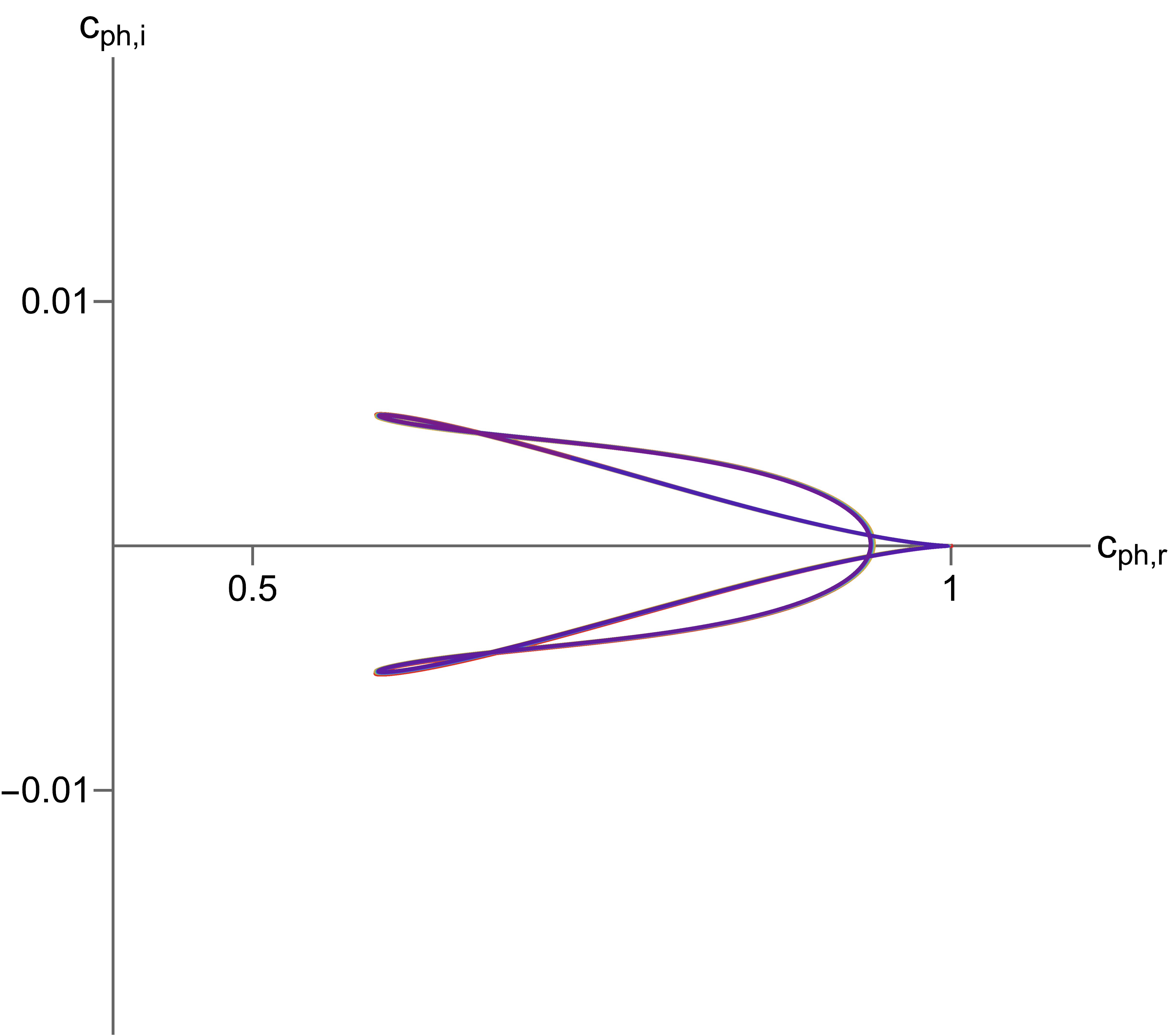}
\caption{
We show a parametric representation of $c_{ph}(t)$ for the continuous time crystalline state, for a large atom number, with $\aleph=100$.
 This corresponds to the black data set of Fig. \ref{ctcthermo}.
    In all cases, we show the first $5$ ms of the correlation function, as indicated in the color bar in Fig. \ref{para}. 
 \label{paralarge}}
\end{figure}

 In Fig. \ref{paralarge}, we display the real and imaginary part of the cavity correlation function $c_{ph}(t)$, for the system parameters described in Sect. \ref{ctcsect}.
  The system has been scaled up via the quantum-classical rescaling utilized in \cite{nowoczyn2025universalquantummeltingquasiperiodic}, with $\aleph=100$.
  The real part of the correlation function  is shown in Fig. \ref{ctcthermo}, as the black data set. 

%

\bibliography{references}

\end{document}